\documentclass[useAMS,fleqn,usenatbib]{mnras}

\usepackage{mathtools}
\usepackage{amssymb}
\usepackage{amsmath}
\usepackage{aas_macros}
\usepackage{graphicx}

\title[Dust Enrichment at $z>6$]{Dust evolution processes in normal galaxies at $z>6$
detected by ALMA}
\author[W.-C. Wang, H. Hirashita, and K.-C. Hou]{
Wei-Chen Wang$^{1,2}$,  Hiroyuki Hirashita$^{1}$ and Kuan-Chou Hou$^{1,3}$\\
$^{1}$Institute of Astronomy and Astrophysics, Academia Sinica, PO Box 23-141, Taipei 10617, Taiwan\\
$^{2}$Department of Mechanical Engineering, National Taiwan University, Taipei 10617, Taiwan\\
$^3$Department of Physics, Institute of Astrophysics, National Taiwan University, Taipei 10617, Taiwan}

\date{\today}

\begin{document}
\maketitle

\begin{abstract}
Recent ALMA observations of high-redshift normal galaxies have been providing a great opportunity to clarify the general origin of dust in the Universe, not biased to very bright special objects even at $z>6$. To clarify what constraint we can get for the dust enrichment in normal galaxies detected by ALMA, we use a theoretical model that includes major processes driving dust evolution in a galaxy; that is, dust condensation in stellar ejecta, dust growth by the accretion of gas-phase metals, and supernova destruction. Using the dust emission fluxes detected in two normal galaxies at $z>6$ by ALMA as a constraint, we can get the range of the time-scales (or efficiencies) of the above mentioned processes. We find that if we assume extremely high condensation efficiency in stellar ejecta ($f_{\mathrm{in}} \ga 0.5$), rapid dust enrichment by stellar sources in the early phase may be enough to explain the observed ALMA flux, unless dust destruction by supernovae in those galaxies is stronger than that in nearby galaxies. If we assume a condensation efficiency expected from theoretical calculations ($f_{\mathrm{in}} \la 0.1$), strong dust growth (even stronger than assumed for nearby galaxies if they are metal-poor galaxies) is required. 
These results indicate that the normal galaxies detected by ALMA at $z>6$ are biased to objects (i) with high dust condensation efficiency in stellar ejecta, (ii) with strong dust growth in very dense molecular clouds, or (iii) with efficient dust growth because of fast metal enrichment up to solar metallicity. A measurement of metallicity is crucial to distinguish among these possibilities.
\end{abstract}


\section{Introduction}
Dust grains in the interstellar medium (ISM) play an important role in radiative and chemical processes.
Dust absorbs
stellar light and re-emits it in the far infrared (FIR), thus affecting the spectral energy distribution (SED) of a galaxy (e.g. \citet{Yajima:2015aa, Schaerer:2015aa}, for recent modeling). Dust also activates $\mathrm{H_2}$ formation since the dust surface serves as a reaction site of H$_2$ formation \citep{Gould:1963ab}. In addition, since dust is composed of metals (silicon, carbon, oxygen, iron, etc.), it traces the metal enrichment in galaxies \citep{Lisenfeld:1998aa, Dwek:1998aa}. 
Therefore, the understanding of dust enrichment in galaxies is of fundamental importance in clarifying the evolutions of galaxies and the interstellar medium.

Galaxies are enriched with metals as a result of their star formation activities. When supernovae (SNe) and asymptotic giant branch (AGB) star winds inject metals in the ISM, a fraction of the metals are condensed into dust \citep{Kozasa:1989aa, Todini:2001aa, Nozawa:2003aa, Bianchi:2007aa,Gall:2011aa}. 
Above a certain metallicity level (typically $\approx 0.1 Z_{\sun}$), grain growth by the accretion of gas phase metals in the dense ISM becomes the dominant source of dust \citep{Dwek:1998aa, Hirashita:1999aa,Inoue:2003aa,Zhukovska:2008aa,Valiante:2011aa,Inoue:2011aa,Mattsson:2012aa,Asano:2013ab}. This process is referred to as `accretion' in this paper. Furthermore, supernova shocks destroy the dust by sputtering \citep{Dwek:1980aa, Nozawa:2006aa}. In this paper, we consider those processes since they dominate the increase or decrease of the dust mass in a galaxy. Indeed, the models including these processes are successful in reproducing the evolution of dust content as a function of metallicity \citep{Kuo:2013aa,Remy-Ruyer:2014aa,de-Bennassuti:2014aa}. However, applications of dust enrichment models have been limited to bright objects at high redshift especially at $z>5$, since it was extremely difficult to detect dust emission from such distant galaxies.

Recent observations by the Atacama Large Millimetre/submillimetre Array (ALMA) have been providing a new window to the studies of origin and evolution of dust in the Universe. Because of its high sensitivity, ALMA has potential for detection of dust emission from high-redshift `normal' galaxies represented by Lyman break galaxies (LBGs) and Lyman $\alpha$ emitters (LAEs) at high redshift ($z\ga 5$) \citep[e.g.][]{Capak:2015aa}. ALMA has expanded the possibility of high-$z$ dust studies, which had been previously limited to very bright objects such as quasars and submillimetre (submm) galaxies. The current frontier of such dust hunting in high-$z$ LBGs and LAEs is $z>6$, where most of the observations ended up with non-detection of dust continuum \citep{Ouchi:2013aa, Ota:2014aa, Maiolino:2015aa, Schaerer:2015aa, Bouwens:2016aa,Aravena:2016aa}.
\citet{Watson:2015aa} recently detected a lensed LBG at $z=7.5$, A1689-zD1 \citep[see also][]{Knudsen:2016aa}. \citet{Willott:2015aa} also reported a firm and a tentative detection of two LBGs at $z\sim 6$. These cases provide important clues to the origin and evolution of dust in the Universe.

\citet{Mancini:2015aa} modelled dust evolution in $z\ga 6$ galaxies. They adopted a cosmological hydrodynamics simulation and apply a dust evolution model that includes dust formation in stellar ejecta, dust growth by accretion, and dust destruction by shocks as post-processing. By assuming a typical accretion time-scale of nearby galaxies (2 Myr at solar metallicity), they compared the calculated redshift evolution dust mass with the observation data of A1689-zD1, finding that the predicted dust mass is too small in comparison with the observation. Thus, they adopted a shorter accretion time-scale as 0.2 Myr to reproduce the observation, concluding that extremely efficient accretion is required. \citet{Michalowski:2015aa} also concluded a necessity of grain growth by accretion for the same object using a much simpler model. Although \citet{Mancini:2015aa}'s model is based on a realistic semi-analytic model, they basically fixed the parameters concerning the dust production and destruction. Since those parameters may be uncertain, it is still worth making an effort of constraining the time-scales and efficiencies of dust production and destruction using A1689-zD1, in order to quantify how fast each process driving dust evolution works. By modeling this object, we can also address the reason for the non-detection of most sources at $z>6$. In other words, we will be able to clarify what kind of dust evolution process determines the detection and non-detection by ALMA. Thus, the predictions in this paper could be useful for the planning and interpretations of future ALMA observations of high-$z$ galaxies.


This paper is organized as follows. In Section 2, we present the dust enrichment model, and explain how we calculate the corresponding dust emission flux. In Section 3, we apply the model to LBGs at $z>6$ detected by ALMA. We discuss our results and parameter dependence in Section 4. In Section 5, we give conclusions on ALMA-detected LBGs. We adopt ($h$, $\Omega_m$, $\Omega_\Lambda$) = (0.7, 0.3, 0.7) for the cosmological parameters.

\section{Method}
The purpose of this paper is to constrain the dust enrichment and destruction processes using stellar and dust luminosities of high-redshift ($z>6$) normal galaxies detected by ALMA. To this aim, we use a simple one-zone chemical evolution model, from which we obtain the evolution of dust mass and stellar mass. Further, we derive the dust temperature by assuming radiative equilibrium of dust. Based on the stellar mass, dust mass, and dust temperature estimated, we further derive the fluxes at rest-frame ultraviolet (UV) and far-infrared (FIR) wavelengths. These fluxes are compared with observational data, which provide constraints to our model. As a consequence, we obtain an insight into the relevant dust enrichment and destruction time-scales (or efficiencies).

\subsection{Dust Enrichment Model}\label{subsec:enrichment}
To put a particular focus on dust formation and destruction, we adopt a simple analytic model in which the galaxy of interest is treated as a single-zone object. We also neglect inflow and outflow for simplicity; that is, the galaxy is treated as a closed box. In reality, dilution of metals and dust would be expected by inflowing gas in the course of galaxy evolution \citep{Feldmann:2015aa}. We will comment on such a dilution effect later in Section \ref{subsec:likely}. The time evolutions of the gas mass ($M_{\mathrm{g}}$), the metal mass ($M_Z$), and the dust mass ($M_{\mathrm{d}}$) are written as \citep{Hirashita:1999aa}
\begin{align}
\frac{dM_{\mathrm{g}}}{dt} &=-\psi + E, \\
\frac{dM_Z}{dt} &=-Z \psi + E_Z, \\
\frac{dM_{\mathrm{d}}}{dt} &=-\mathcal{D} \psi +f_{\mathrm{in}} E_Z -\frac{M_{\mathrm{d}}}{\tau_{\mathrm{SN}}} +\frac{M_{\mathrm{d}}(1-f_Z)}{\tau_{\mathrm{acc}}},
\label{eq:dMd_dt}
\end{align}
where $\psi$ is the star formation rate (SFR), $E$ is the total ejection rate of stellar ejecta, $E_Z$ is the total ejection rate of metals in stellar ejecta, $Z$ is the metallicity (i.e.\ $Z \equiv M_Z/M_{\mathrm{g}}$), $\mathcal{D}$ is the dust-to-gas ratio (i.e.\ $\mathcal{D} \equiv M_{\mathrm{d}}/M_{\mathrm{g}}$), $f_{\mathrm{in}}$ is the condensation efficiency of metals in stellar ejecta, $f_Z$ is the mass fraction of metals locked in the dust (i.e.\ $f_Z \equiv M_{\mathrm{d}}/M_Z$), $\tau_{\mathrm{acc}}$ is the time-scale of dust growth by accretion, and $\tau_{\mathrm{SN}}$ is the time-scale of dust destruction by SN shocks.

For the convenience of analytic treatment, we adopt the instantaneous recycling approximation \citep{Tinsley:1980aa}.
The mass ejection rate of stellar ejecta $E$ and that of metals $E_Z$ are both proportional to the star formation rate as $E=\mathcal{R} \psi$, and $E_Z=\mathcal{Y}_Z \psi$, where $\mathcal{R}$ and $\mathcal{Y}_Z$ are the returned fraction and metal yield of stars, respectively.
We evaluate $\mathrm{R}$ and $\mathcal{Y}_Z$ based on equations (8) and (10) in \citet{Inoue:2011aa} and the fitting formulae of remnant mass $w(m,\, Z)$ and metal yield $m_Z(m)$ from the same reference, using $5 M_{\sun}$ as the turn-off mass. Consequently we obtain $\mathcal{R}=0.1644$ and $\mathcal{Y}_Z=0.0139$.
These values are not sensitive to the turn-off mass chosen \citep{Hirashita:2011aa}, and the results below are far more sensitive to the parameters directly related to the dust evolution (i.e.\ $f_\mathrm{in}$, $\tau_\mathrm{SN}$, and $\tau_\mathrm{acc}$).

The SFR is assumed to be regulated by a non-changing star formation time-scale, $\tau_{\mathrm{SF}}$, as $\psi \equiv M_{\mathrm{g}}/\tau_{\mathrm{SF}}$. Since the efficiency of dust growth by accretion is proportional to the metallicity, the accretion time-scale is written by introducing the dust growth time-scale at solar metallicity, $\tau_\mathrm{acc,0}$ as \citep{Hirashita:2011aa, Mancini:2015aa}
\begin{equation}
\tau_{\mathrm{acc}}=\tau_{\mathrm{acc,0}}\left( \frac{Z}{\mathrm{Z}_{\sun}} \right)^{-1}.
\end{equation}
Therefore, under a given condensation efficiency $f_\mathrm{in}$, the dust evolution is determined by a given time-scale parameter set ($\tau_{\mathrm{SF}}$, $\tau_{\mathrm{acc,0}}$, $\tau_{\mathrm{SN}}$). It is reasonable that the dust accretion time-scale and SN destruction time-scale are scaled with $\tau_{\mathrm{SF}}$ for the following reasons. SNe occur more often if more stars formed \citep{Dwek:1980aa}. More accretion happens if more gas is contained in the dense ISM phase, which also hosts star formation \citep{Hirashita:2011aa}. Therefore, we normalize the accretion and SN destruction time-scales to the star formation time-scale by introducing two parameters, $\zeta_{\mathrm{acc,0}}$ and $\zeta_{\mathrm{SN}}$, as
\begin{align}
\tau_{\mathrm{acc,0}}&=\zeta_{\mathrm{acc,0}}\tau_{\mathrm{SF}}, \\
\tau_{\mathrm{SN}}&=\zeta_{\mathrm{SN}}\tau_{\mathrm{SF}}.
\end{align}
Furthermore, if $t$ is normalized to $\tau_{\mathrm{SF}}$, the same time evolution of $(M_{\mathrm{g}}$, $M_Z$, $M_{\mathrm{d}})$ is obtained as a function of $t/\tau_{\mathrm{SF}}$, for the same values of $f_\mathrm{in}$, $\zeta_{\mathrm{acc,0}}$ and $\zeta_{\mathrm{SN}}$.

For the initial condition, we adopt $M_{\mathrm{g}}=M_0$, and $M_Z=M_{\mathrm{d}}=0$. Since we adopt a closed-box model, the star mass $M_{\star}$ is estimated as
\begin{equation}
M_{\star}(t)=M_0-M_{\mathrm{g}}(t).
\end{equation}
Note that all the masses mentioned above are proportional to the initial total mass  $M_0=M_{\mathrm{g}}(t=0)$.

\subsection{FIR and UV luminosity}
\citet{Schaerer:2015aa} derived the following relation between UV magnitude at rest 1500 \AA, $M_{1500}$, and stellar mass $M_{\star}$, from SED fitting to a sample of LBGs, including nebular emission and dust attenuation:
\begin{equation}
\log \left( \frac{M_{\star}}{M_{\sun}} \right)=-0.45\times(M_{1500}+20)+9.11.
\label{eq:Mstar_M1500}
\end{equation}
This gives a relation between $M_\star$ calculated by our model and
$M_{1500}$.
We convert $M_{1500}$ to the UV luminosity $L_{\mathrm{UV}}$ by adopting
\begin{equation}
M_{1500}= -2.5 \log \left( \frac{L_{\mathrm{UV}}}{L_{\sun}} \right)+5.73,
\label{eq:M1500_LUV}
\end{equation}
where the reference values of $M_{1500}$ and $L_{\mathrm{UV}}$ are taken from Himiko's ones \citep{Ouchi:2013aa}. As a consequence, we get $L_\mathrm{UV}$  using $M_{\star}$ calculated in our dust evolution model in Section 2.1.
Note that $L_{\mathrm{UV}}$ is almost proportional to $M_{\star}$. The estimated $L_{\mathrm{UV}}$ already includes the effect of attenuation, but the following conclusions are not affected by the treatment of attenuation because of the small attenuation of our sample.

Although the stellar mass is estimated more precisely using data at longer wavelengths, we use UV magnitude for the following reasons. Since LBGs are UV-selected galaxies, they always have data in the rest UV, while the availability of longer-wavelength data depends on how extensively those objects are followed up. More importantly, the above formula is already tested for a sample of LBGs by \citet{Schaerer:2015aa}. Their formula is derived by SED fitting including nebular emission to a sample of LBGs at $z \sim 7$, which means that their formula is applicable to the galaxies of interest in our present paper. Therefore, we take advantage of their convenient expression to make our formulation analytically manageable. In addition, we compare the stellar masses estimated by the spectral fitting in \citet{Watson:2015aa} and by the formula in \citet{Schaerer:2015aa}, and find that they differ by a factor of 1.4. This is small enough, and does not affect our conclusions.

Besides, the dust temperature ($T_\mathrm{d}$) necessary to estimate the submm flux is derived from the condition of radiative equilibrium. Assuming that the UV radiation from stars is the dominant heating source of dust in star-forming galaxies (e.g.\ Buat \& Xu 1996), the radiative equilibrium (the balance between the energies absorbed and emitted by a dust grain) is written as \citep{Hirashita:2014aa}
\begin{equation}
\pi a^2 Q_{\mathrm{UV}}\frac{L_{\mathrm{UV}}}{4\pi R^2}
= \int_0^{\infty} 4\pi\kappa_{\nu}\left(\frac{4}{3}\pi a^3s\right) B_{\nu}(T_{\mathrm{d}})d\nu,
\label{eq:rad_eq}
\end{equation}
where $\kappa_{\nu}$ is the dust mass absorption coefficient, $s$ is the dust material density, $a$ is the dust grain radius, $Q_{\mathrm{UV}}$ is the ratio of the absorption cross-section to the geometric cross-section for UV radiation (which we assume to be $1$), and $R$ is the radius of the spatial dust distribution. We adopt a power-law with an index of $\beta$ for the mass absorption coefficient, $\kappa_{\nu}=\kappa_{158} (\nu/\nu_{158})^{\beta}~\mathrm{cm}^2~\mathrm{g}^{-1}$ following \citep{Hirashita:2014aa}, where $\kappa_{158}$ is the dust mass absorption coefficient at wavelength $\lambda=158~\mu$m, and $\nu_{158}=1.9$ THz is the frequency corresponding to $\lambda=158~\mu$m. Among the grain species in \citet{Hirashita:2014aa}, we adopt graphite:  $\kappa_{158}=20.9$ $\mathrm{cm}^2$ $\mathrm{g}^{-1}$, $\beta=2$, and $s=2.26$ $\mathrm{g}$ $\mathrm{cm}^{-3}$ with grain radius $a=0.1$ $\mu$m. We discuss the dependence on grain properties in Section \ref{subsec:grain_properties}.

Since our selected galaxies are located at high redshifts, we also consider heating by the cosmic microwave background (CMB). We use the following equation to correct the dust temperature for the CMB heating at redshift $z$ \citep{da-Cunha:2013aa}:
\begin{equation}
T_{\mathrm{d}}(z)=\left\{ (T_{\mathrm{d}}^{z=0})^{4+\beta }+(T_{\mathrm{CMB}}^{z=0})^{4+\beta}\left[ (1+z)^{4+\beta}-1\right]\right\}^{\frac{1}{4+\beta}},
\end{equation}
where $T_{\mathrm{CMB}}^{z=0}=2.7$ K is the CMB temperature at $z=0$ and $T_{\mathrm{d}}^{z=0}$ is the dust temperature calculated in equation (\ref{eq:rad_eq}).
We hereafter denote the corrected dust temperature $T_\mathrm{d}(z)$ as $T_\mathrm{d}$.

Now, using the dust mass $M_{\mathrm{d}}$ calculated in Section \ref{subsec:enrichment} and the dust temperature $T_{\mathrm{d}}$ obtained above, we can estimate the dust emission flux ($f_{\mathrm{d},\nu}$) at observational frequency $\nu$ as (e.g.\citet{Dayal:2010aa})
\begin{equation}
f_{\mathrm{d},\nu}=\frac{(1+z)\kappa_{(1+z)\nu}M_{\mathrm{d}}B_{(1+z)\nu}(T_{\mathrm{d}})}{d_L^2},
\label{eq:flux_dust}
\end{equation}
where $d_L$ is the luminosity distance \citep{Carroll:1992aa} of the object.
We also estimate the UV flux $f_{\mathrm{UV}}$ by the following equation:
\begin{equation}
f_{\mathrm{UV}}=\frac{L_{\mathrm{UV}}}{4\pi d_L^2}.
\end{equation}
We can see that $f_{\mathrm{d},\nu}$ is proportional to $M_{\mathrm{d}}$ from equation (\ref{eq:flux_dust}). As mentioned in Section \ref{subsec:enrichment}, $M_{\mathrm{d}}$ and $M_{\star}$ are proportional to $M_0$. Moreover, as shown above, $L_{\mathrm{UV}}$ is almost proportional to $M_{\star}$. This indicates that both $f_{\mathrm{d},\nu}$ and $f_{\mathrm{UV}}$ are (almost) proportional to $M_0$. Thus, to effectively eliminate the dependence on the unknown $M_0$, we compare $\nu f_{\mathrm{d},\nu}/f_{\mathrm{UV}}$ with observations. We multiply $\nu$ to $f_{\mathrm{d},\nu}$ to make $\nu f_{\mathrm{d},\nu}/ f_{\mathrm{UV}}$ dimensionless. Since we adopt frequencies near to the SED peak, $\nu f_{\mathrm{d},\nu}$ is roughly the total dust emission flux. Also, since a slight dependence on $M_0$ still remains for $f_{\mathrm{UV}}$ (note that $L_{\mathrm{UV}}$ is not perfectly proportional to $M_{\star}$; see equations \ref{eq:Mstar_M1500} and \ref{eq:M1500_LUV}), we adopt a specific value of $M_0=3\times10^{10}M_{\sun}$ although this choice does not affect the following discussions.

\subsection{Observational Data}
We compare the calculated $\nu f_{\mathrm{d},\nu}/f_{\mathrm{UV}}$ with observations.
As mentioned in the Introduction, we chose normal galaxies at $z>6$
(i.e.\ the current frontier for dust hunting of normal galaxies) identified in the optical
with the dust
continuum detected. This sample provides us with opportunity to access `normal' galaxies
with moderate SFRs $\sim 10$ M$_{\sun}$ yr$^{-1}$, and enables us to derive a more general picture
of dust enrichment in the earliest Universe accessible by ALMA, not biased to submm galaxies
or QSOs \citep{Mancini:2015aa}. So far, A1689-zD1 \citep{Watson:2015aa,Knudsen:2016aa} and
CLM1 \citep{Willott:2015aa} satisfy this criterion. A1698-zD1 is strongly lensed by
a factor of 9.5 and the redshift (7.5) is determined by the spectroscopic break feature
at the restframe UV
wavelengths. The dust continuum of CLM1 at $z=6.2$ was tentatively detected
at the 2$\sigma$ level by ALMA. Because of the rareness of the sample, we use this flux
as detected, keeping in mind that this might overestimate the dust emission (or the efficiency
of dust enrichment) in our model. \citet{Willott:2015aa} also reported a more significant detection
of dust continuum emission for WMH 5; however, the emitting regions of dust continuum and
stars are clearly different in this object, indicating that this object is a merging system
between a dusty galaxy and a dust-free galaxy. Therefore, the UV radiation observed
is not likely to be the heating source of the dust in this object, which means that our
model cannot be applied to it. In Table \ref{tab:sample}, we list the observational data adopted in this paper.

\begin{table*}
\begin{center}
\begin{minipage}{100mm}
\caption{Galaxy sample at $z>6$ adopted.}
\label{tab:sample}
\begin{tabular}{@{}lcccccc}
\hline
Galaxies & redshift & $\nu (1+z)$ & $f_{\mathrm{d},\nu}$ & $R$  & $L_{\mathrm{UV}}$  & Ref. $^a$ \\
& & (Hz) & (mJy) & (kpc) & ($L_{\sun}$) & \\
\hline
A1689-zD1 & $7.5\pm0.2$ & $1.9\times10^{12}$ & $0.066\pm0.013$ $^b$ & $0.7$ & $1.8\times10^{10}$ $^b$ & 1\\
CLM1 & $6.17\pm0.0003$ & $1.9\times10^{12}$ & $0.044\pm0.026$ $^c$ & $2.65$ & $3.61\times10^{10}$ & 2\\
\hline
\end{tabular}
\\
\textit{Note.}\\
$^a$ References 1) Watson et al. (2015). 2) Willott et al. (2015). \\
$^b$ Corrected for the gravitational lensing, magnification by a factor 9.3. \\
$^c$ Neglected the low gravitational magnification 1.13. \\
\end{minipage}
\end{center}
\end{table*}

\section{Results}

\subsection{Overall behaviour}
\begin{figure}
\includegraphics[width=0.4\textwidth]{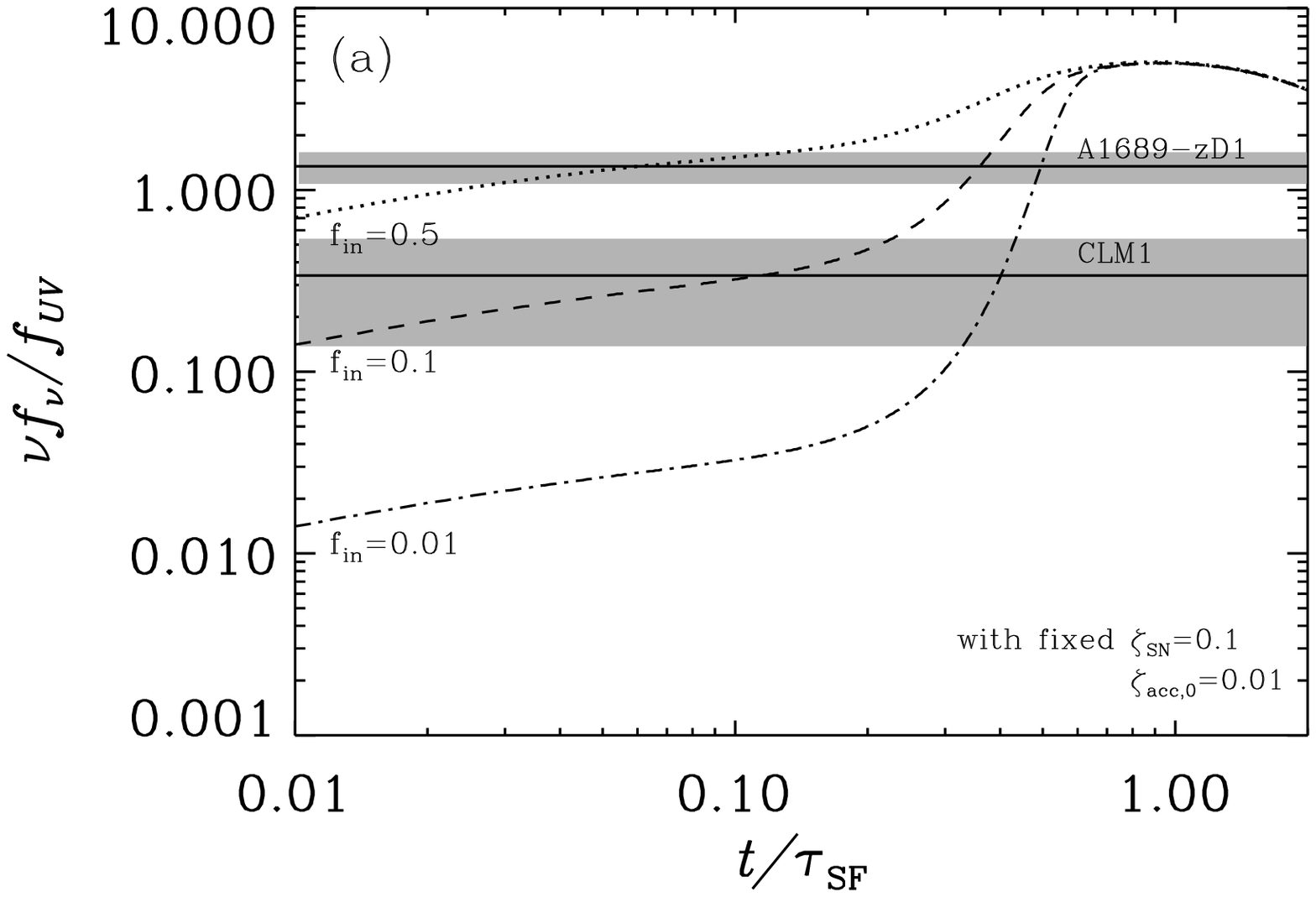}
\includegraphics[width=0.445\textwidth]{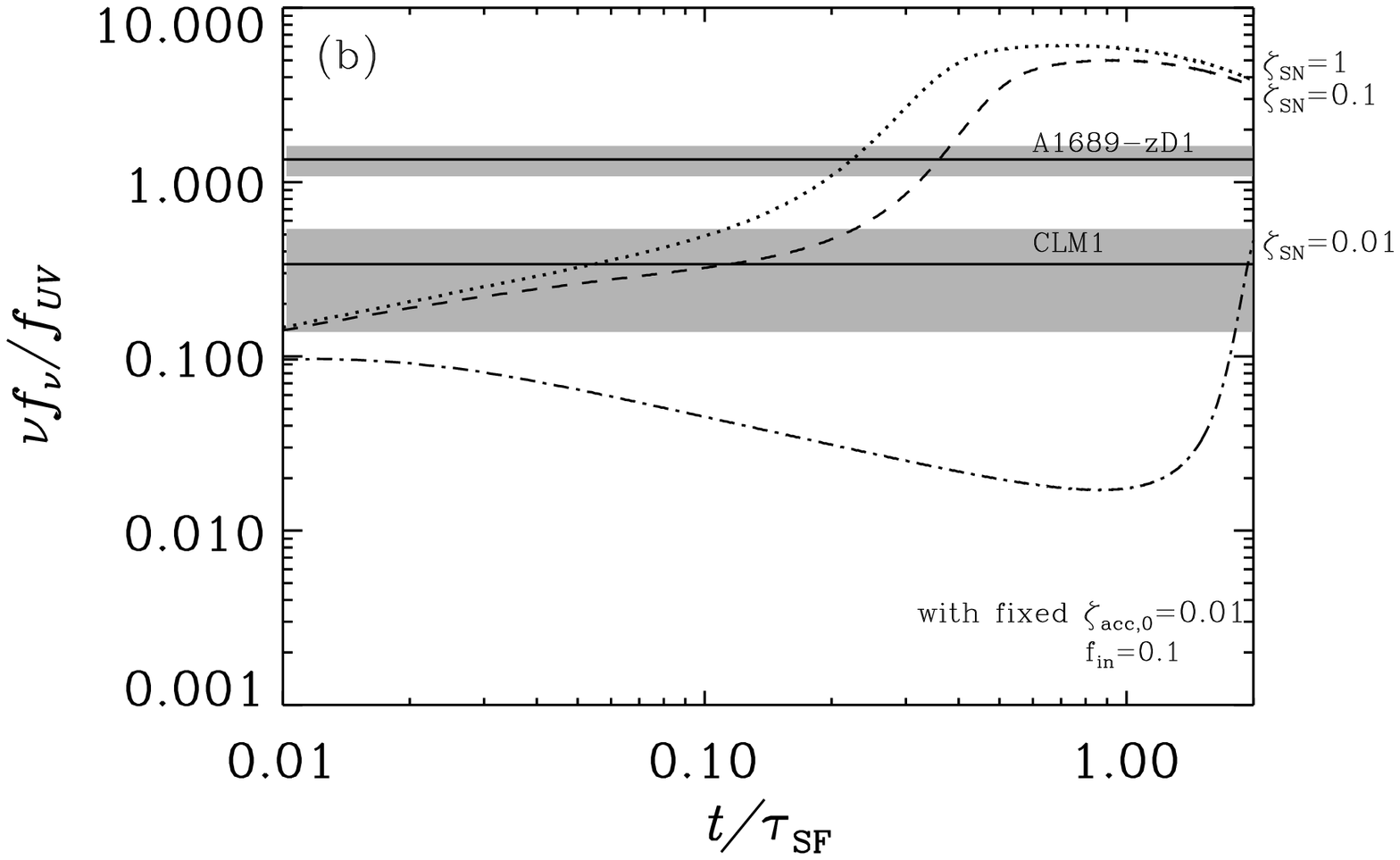}
\includegraphics[width=0.455\textwidth]{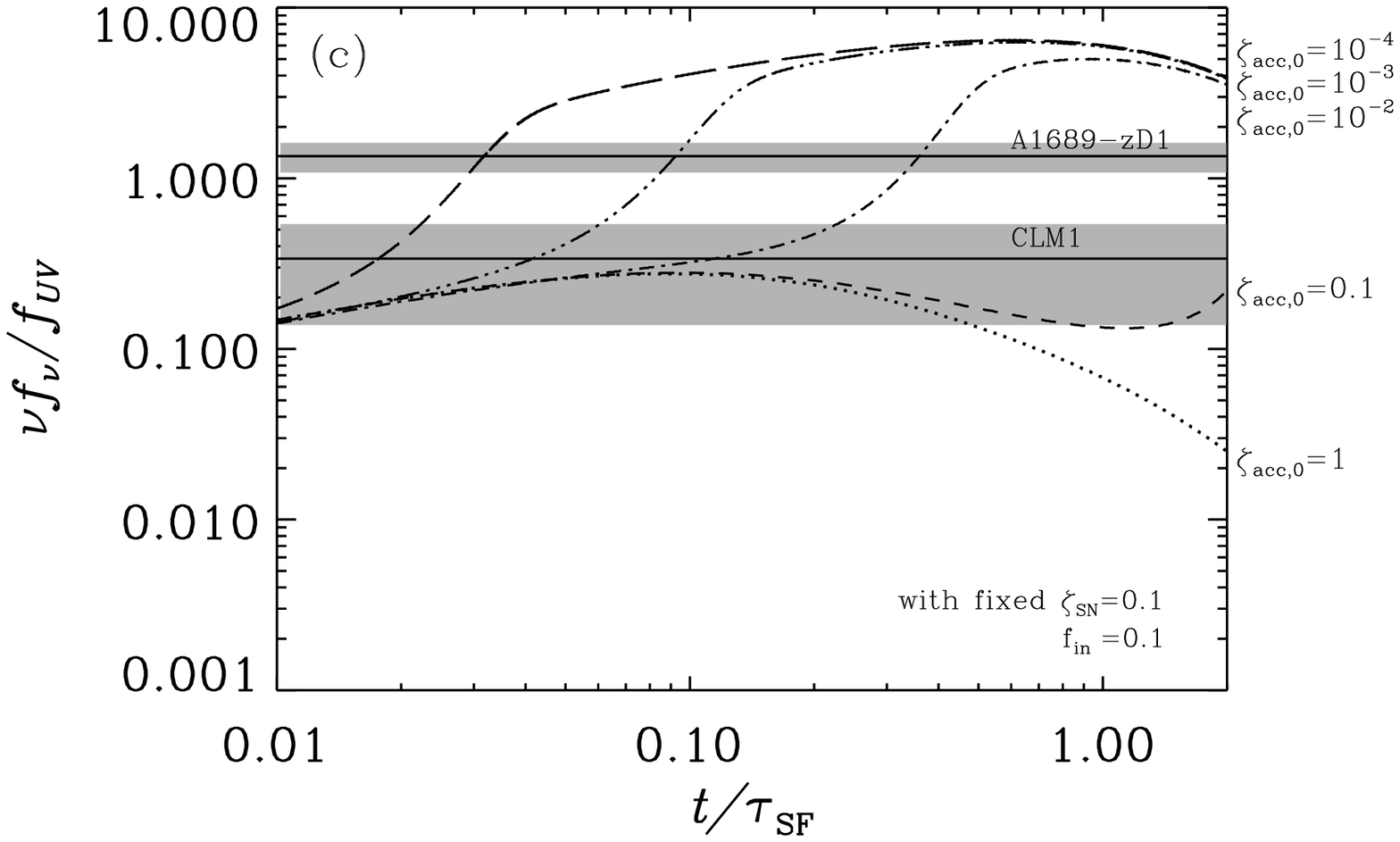}
\includegraphics[width=0.4\textwidth]{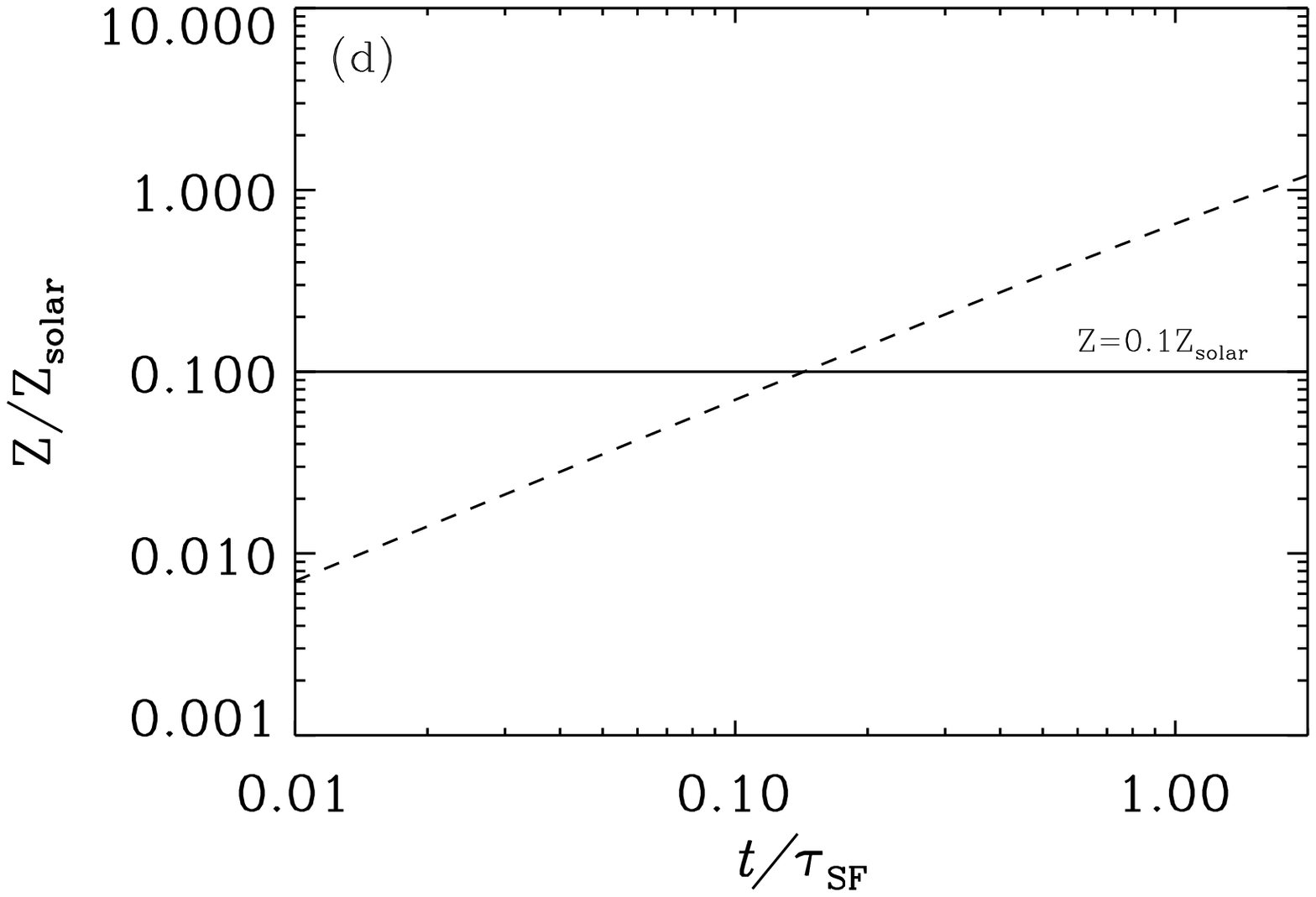}
\caption{
Normalized dust emission flux $\nu f_{\nu}/f_{\mathrm{UV}}$ as a function of the normalized time $t/\tau_\mathrm{SF}$. The horizontal lines at $\nu f_{\nu}/f_{\mathrm{UV}}=1.35$ and $0.34$ are the observed values of A1689-zD1 and CLM1, respectively. The shaded area shows the error of each object. (a) The dotted, dashed, and dotted-dashed lines show the results for $f_{\mathrm{in}}=0.5$, $0.1$, and $0.01$, respectively, with $\zeta_{\mathrm{SN}}=0.1$ and $\zeta_{\mathrm{acc,0}}=0.01$. (b) The dotted, dashed, and dotted-dashed lines are for $\zeta_{\mathrm{SN}}=1$, $0.1$ and $0.01$, respectively, with $\zeta_{\mathrm{acc,0}}=0.01$ and $f_{\mathrm{in}}=0.1$. (c) The long-dashed, tripple-dotted-dashed, dotted-dashed, dashed and dotted lines are for $\zeta_{\mathrm{acc,0}}=10^{-4}$, $10^{-3}$, $0.01$, $0.1$, and $1$, respectively with $\zeta_{\mathrm{SN}}=0.1$ and $f_{\mathrm{in}}=0.1$. (d) Metallicity evolution (dotted line). The horizontal solid line shows $Z=0.1$ Z$_{\sun}$: LGBs at $z>6$ typically have metallicities below this level \citep{Mancini:2015aa}.}
\label{fig:basic_ev}
\end{figure}

In Fig.\ \ref{fig:basic_ev}, we plot the normalized flux $\nu f_{\mathrm{d},\nu}/f_{\mathrm{UV}}$ versus the normalized time $t/\tau_{\mathrm{SF}}$. To separate the effect of each process, we fix two parameters in the set ($f_{\mathrm{in}}$, $\zeta_{\mathrm{acc,0}}$, $\zeta_{\mathrm{SN}}$) and change the other in each panel. In Fig.\ \ref{fig:basic_ev}a, we observe that the effect of $f_{\mathrm{in}}$ appears at the early stage and that all the cases converge to the same evolution at later times. Thus, we can conclude that dust condensation in stellar ejecta has an influence on the early evolution. This is consistent with the conclusion by \citet{Hirashita:1999aa} that the stellar dust condensation dominates the dust abundance in the early (low-metallicity) stage of galaxy evolution. Besides, from equation (\ref{eq:dMd_dt}), we find that dust growth by accretion and SN destruction are dependent on $M_{\mathrm{d}}$. Thus, as the dust abundance becomes large, the evolution is determined by these two processes; this is why all the cases with different $f_\mathrm{in}$ converge to the same evolution. 

In Figs.\ \ref{fig:basic_ev}b and c, we show the effect of changing $\zeta_{\mathrm{SN}}$ and $\zeta_{\mathrm{acc,0}}$, respectively. In contrast to the cases with different $f_\mathrm{in}$, the effects of the variations of these two parameters are prominent at a late stage of evolution as expected from the discussion in the previous paragraph. SN destruction decreases and accretion increase the dust emission relative to the stellar emission as expected. Thus, we conclude that SN destruction and dust accretion are important when the system is enriched with dust and metals. We will characterize the metallicity level at which these processes, especially accretion, become dominant in Section~\ref{subsec:metallicity}.

Metallicity is often used as a measure for the evolutionary stage of a
galaxy. In Fig.\ \ref{fig:basic_ev}, we show the evolution of metallicity
as a function of $t/\tau_\mathrm{SF}$. We observe that the metallicity
reaches 0.1 Z$_{\sun}$ at $t/\tau_\mathrm{SF}=0.15$.
According to the theoretical calculations by \citet{Mancini:2015aa}, the metallicity of the LBGs
is less than 0.1 Z$_{\sun}$. Therefore, we first put a constraint on
the metallicity as $Z<0.1$ Z$_{\sun}$. Since the galaxies detected by
ALMA may be biased to evolved ones as we discuss later, we also
examine a case where we do not put any constraint on the metallicity.

\subsection{Constraint on the dust evolution parameters}
The dust evolution parameters $(f_\mathrm{in},\,\zeta_\mathrm{acc,0},\,\zeta_\mathrm{SN})$ are constrained by the condition that the observed $\nu f_{\nu}/f_{\mathrm{UV}}$ is achieved. This condition is written as
\begin{equation}
\text{max(} \nu f_{\nu}/f_{\mathrm{UV}} \text{)}_{\mathrm{model}} \geq  (\nu f_{\nu}/f_{\mathrm{UV}})_{\mathrm{obs}},\label{eq:cond}
\end{equation}
where the subscripts `model' and `obs' denote the calculated and observed values, respectively, and max is the maximum value over the galaxy lifetime.

Before analyzing the result, we first discuss the probable ranges of the three parameters $(f_\mathrm{in},\,\zeta_\mathrm{acc,0},\,\zeta_\mathrm{SN})$. As compiled in Kuo et al. (2013), the value of the condensation efficiency $f_{\mathrm{in}}$ expected from theoretical calculations has large uncertainty. Thus, we investigate the value from 0.01 to 0.5 with $f_\mathrm{in}=0.1$ being the fiducial value. Also, \citet{McKee:1989aa} and \citet{Lisenfeld:1998aa} show that $\beta_{\mathrm{SN}}=M_{\mathrm{g}}/(\tau_{\mathrm{SN}} \psi)=\tau_{\mathrm{SF}}/\tau_{\mathrm{SN}}$ (which is equivalent with $1/\zeta_{\mathrm{SN}}$ in our paper) is $\approx 10$. For a wide parameter survey, we investigate the range $0.01 \la \zeta_{\mathrm{SN}} \la 1$. We are interested in as efficient accretion as suggested by \citet{Mancini:2015aa}. To consider such efficient accretion, we focus on $\zeta_{\mathrm{acc,0}}$ down to $10^{-4}$ (see also the discussion in Section \ref{subsec:likely}).

\begin{figure}
\begin{center}
\includegraphics[width=0.45\textwidth]{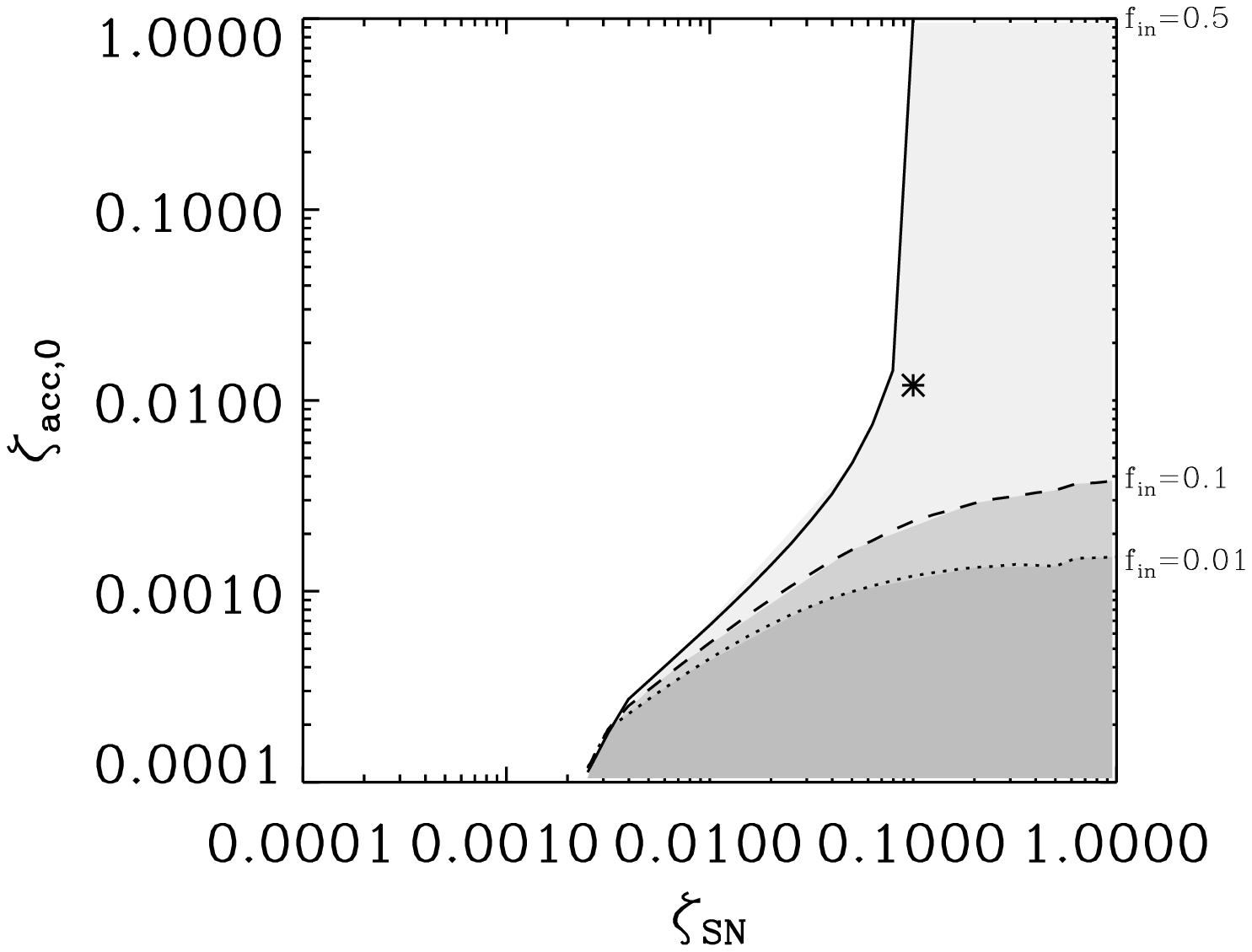}
\includegraphics[width=0.45\textwidth]{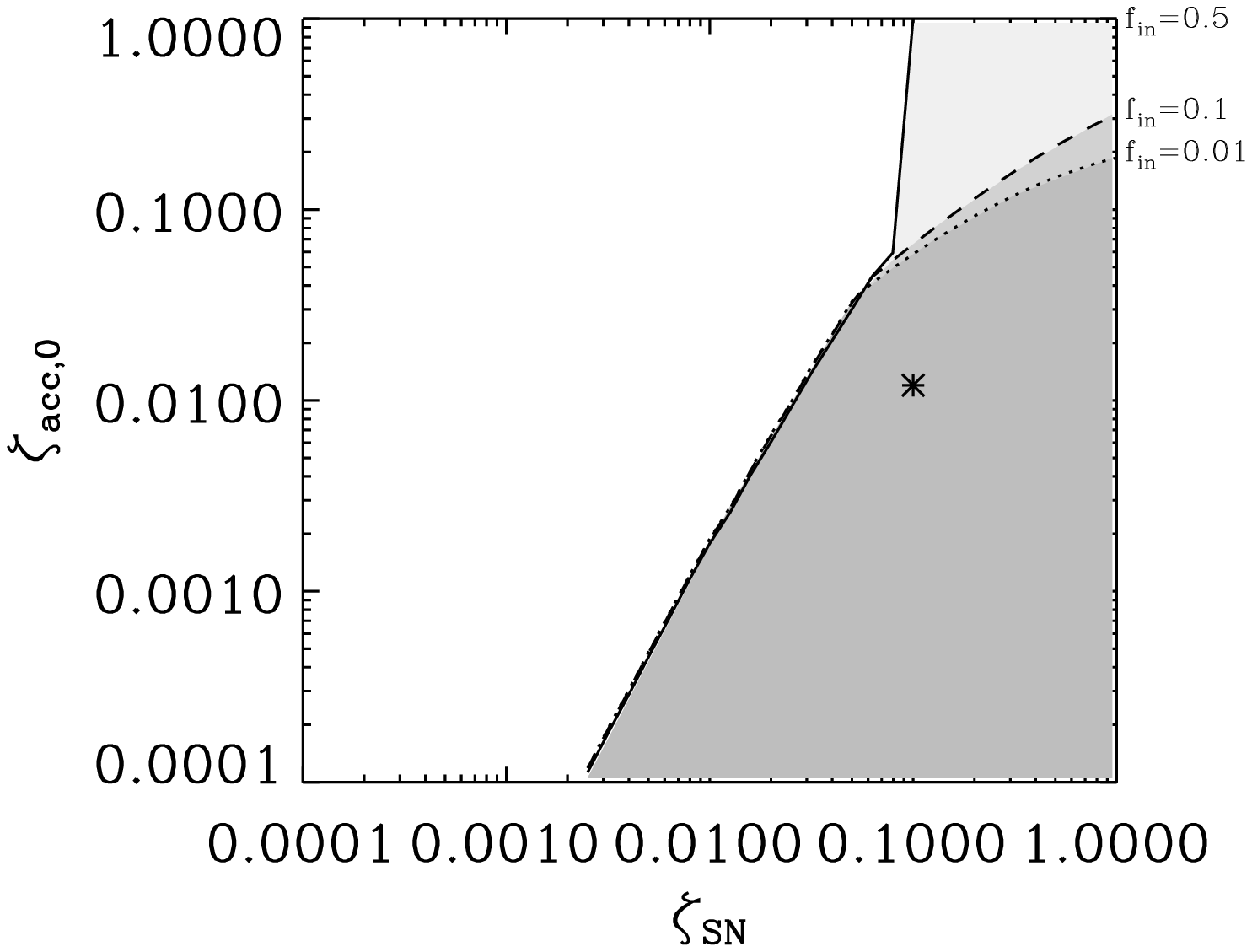}
\caption{Ranges of the time-scale parameters for accretion and SN destruction $(\zeta_\mathrm{acc,0},\,\zeta_\mathrm{SN})$ constrained by A1689-zD1 for $f_\mathrm{in}=0.5$, 0.1, and 0.01. The region on the right and lower side (shaded region) of each line satisfies the criterion. The solid, dashed, and dotted lines are the boundary lines for $f_\mathrm{in}=0.5$, 0.1, and 0.01, respectively. The upper and lower panel are with and without the criterion of metallicity $Z>0.1$ Z$_{\sun}$, respectively. The asterisk refers to the standard accretion and destruction values of nearby galaxies,
$(\zeta_\mathrm{acc,0},\,\zeta_\mathrm{SN})=(0.012,\, 0.1)$}
\label{fig:constraint}
\end{center}
\end{figure}

In the upper panel of Fig.\ \ref{fig:constraint}, we show the area on the ($\zeta_{\mathrm{acc,0}}$, $\zeta_{\mathrm{SN}}$) plane in which the set of ($f_{\mathrm{in}}$, $\zeta_{\mathrm{acc,0}}$, $\zeta_{\mathrm{SN}}$) of A1689-zD1 satisfies the above condition (equation \ref{eq:cond}) at $Z<0.1$ Z$_{\sun}$ (expected metallicity for LBGs by \citealt{Mancini:2015aa}). We find that for $f_{\mathrm{in}} \lesssim 0.1$, $\zeta_{\mathrm{acc,0}}$ much smaller than $\zeta_{\mathrm{SN}}$ is necessary to explain the observational data. This is because only when the accretion time-scale is shorter than the supernova time-scale, the dust could have chance to grow before being destroyed by SNe. If we assume a standard value of $\zeta_\mathrm{SN}\sim 0.1$ as suggested for nearby galaxies \citep{McKee:1989aa}, small values of $\zeta_{\mathrm{acc,0}}$ ($< 0.004$) are required. For large $f_{\mathrm{in}} > 0.5$, however, the constraint on $\zeta_{\mathrm{acc,0}}$ is not stringent since the dust produced by stellar sources makes it easier to increase $\nu f_{\mathrm{d},\nu}/f_{\mathrm{UV}}$ up to the observed value regardless of accretion. Nevertheless, even for $f_{\mathrm{in}} = 0.5$, we need $\zeta_{\mathrm{acc,0}}<0.01$ if SN destruction is efficient ($\zeta_{\mathrm{SN}} \lesssim 0.05 $). As we discuss later, theoretical dust condensation calculations indicate that $f_{\mathrm{in}}$ is likely to be less than 0.5, which means that dust growth by accretion is important to explain the dust emissions detected by ALMA for A1689-zD1.

Next we remove the condition for the metallicity in the lower panel of Fig.\ \ref{fig:constraint} by allowing a possibility that LBGs detected by ALMA are more metal-rich than expected (i.e.\ $Z>0.1$ Z$_{\sun}$). We can see that the trend of $f_{\mathrm{in}}=0.5$ barely changes, meaning that the large efficiency of stelar dust production produces a sufficient amount of dust in the early phase of evolution as shown in Fig. 1a. In contrast, the condition for $\zeta_\mathrm{acc,0}$ is much more relaxed  for $f_\mathrm{in}=0.01$ and 0.1. This means that allowing for a metal-rich stage is essential for accretion, since the efficiency of accretion is proportional to the metallicity. As shown later, $\zeta_\mathrm{acc,0}\sim 0.1$ is similar to the value inferred for nearby galaxies. Thus, if the LBGs detected by ALMA at $z>6$ are solar-metallicity objects, their rich dust content can be explained by assuming an accretion efficiency expected for nearby galaxies.

\begin{figure}
\begin{center}
\includegraphics[width=0.45\textwidth]{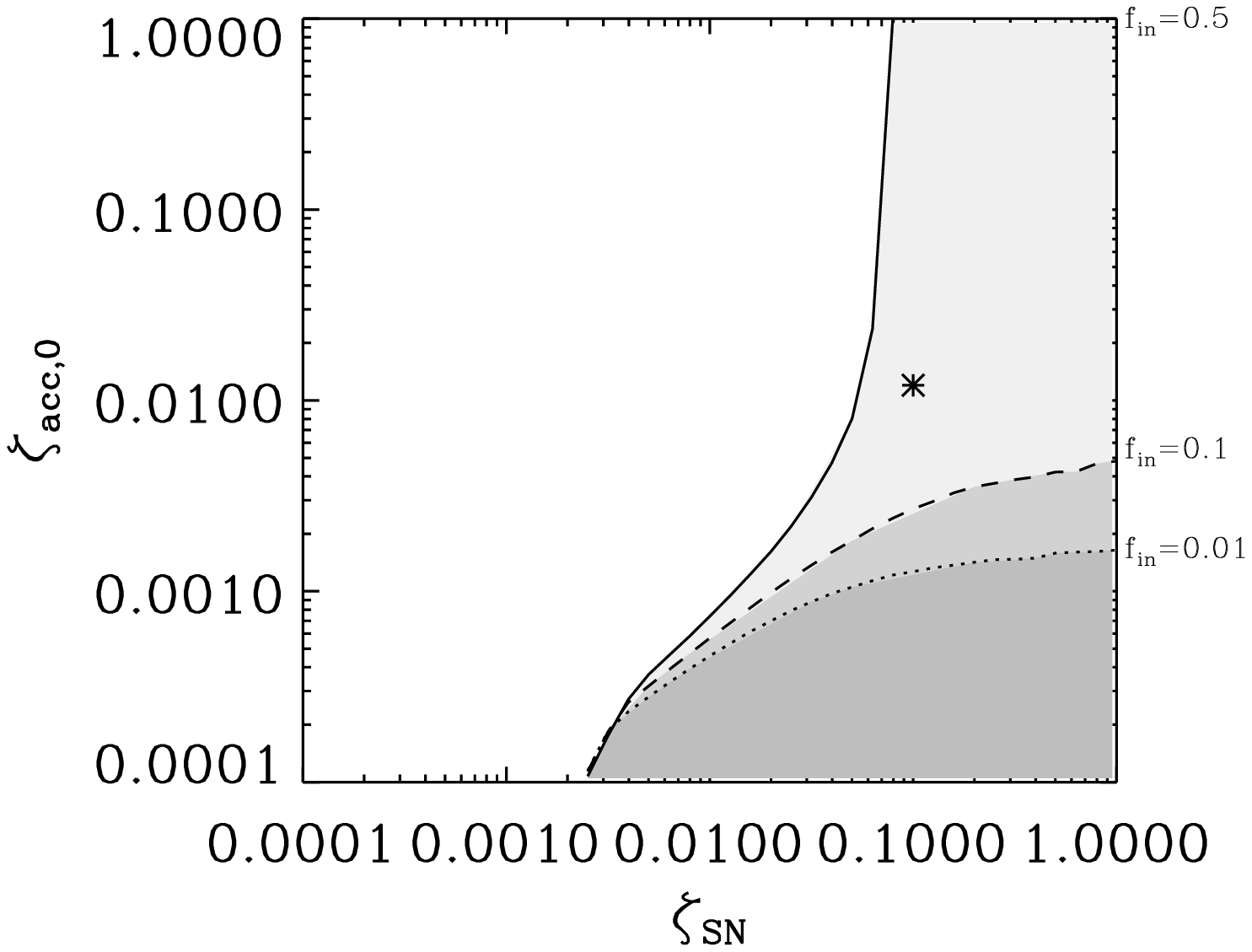}
\includegraphics[width=0.45\textwidth]{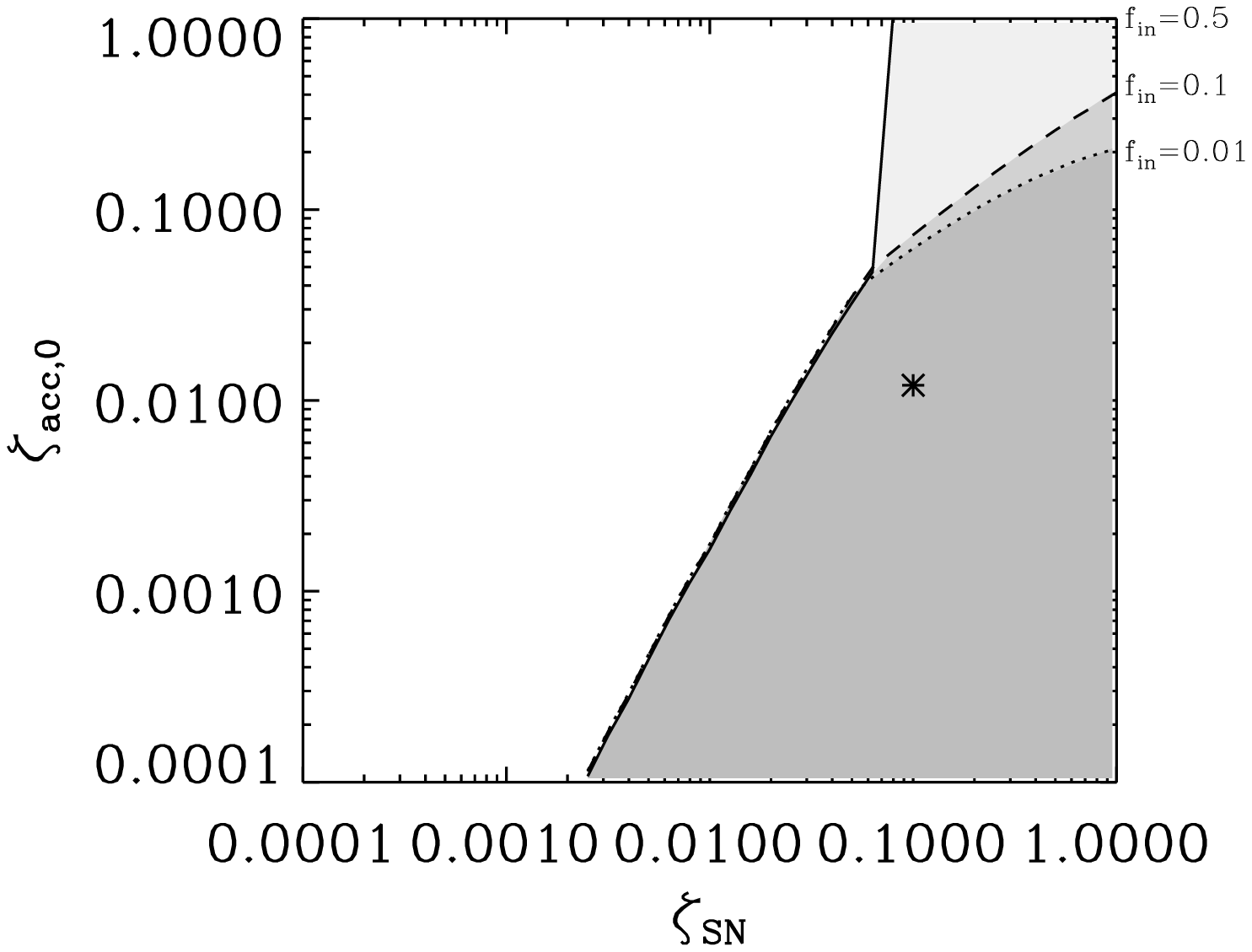}
\caption{Same as Fig.\ \ref{fig:constraint} but for the other object, CLM1.}
\label{fig:constraint-2}
\end{center}
\end{figure}

The upper and lower panel of Fig.\ \ref{fig:constraint-2} show the results for CLM1. Comparing Figs.\ \ref{fig:constraint} and \ref{fig:constraint-2}, we observe that the condition for dust formation and SN destruction is similar for both objects. That is, for small dust condensation efficiency in stellar ejecta ($f_\mathrm{in}\la 0.1$), both object need very efficient dust accretion ($\zeta_\mathrm{acc,0}\la 0.004$) under the metallicity constraint $Z<0.1$ Z$_{\sun}$. This constraint on the accretion time-scale is much relaxed if we allow the metallicity to be higher than 0.1 Z$_{\sun}$.
The fact that we obtain very similar constraints for both objects indicates that detection of LGBs by ALMA requires a similar range for $(f_\mathrm{in},\,\zeta_\mathrm{acc,0},\,\zeta_\mathrm{SN})$.

\section{Discussion}

\subsection{Likely values for the parameters}\label{subsec:likely}

\citet{Knudsen:2016aa} derived the SFR of A1689-zD1 as $12$ M$_{\sun}$ yr$^{-1}$ from the submm continuum. Although the gas has not been detected, \citet{Watson:2015aa} derived a gas mass as $2\times 10^9$ M$_{\sun}$ based on the Schmidt-Kennicutt law, which is equivalent to assuming a star formation time-scale of $1.7\times 10^8$ yr. \citet{Mancini:2015aa} showed, using their semi-analytic models of galaxy formation and evolution, that A1689-zD1 requires very short accretion time-scale, $\tau_{\mathrm{acc,0}}=0.2$ Myr. Thus, the literature value indicates that $\zeta_{\mathrm{acc,0}}\sim 1.2\times 10^{-3}$. Our constraint obtained in Fig.\ \ref{fig:constraint} shows $\zeta_\mathrm{acc,0}\la 4\times 10^{-3}$ if we impose the metallicity constraint $Z<0.1$ Z$_{\sun}$ as indicated by \citet{Mancini:2015aa}. Thus, we confirm their conclusion about the necessity of strong accretion quantitatively \citep[see also][]{Michalowski:2015aa}. Such a high accretion efficiency is realized if the molecular clouds are very dense \citep{Kuo:2013aa,Schneider:2016aa}. However, we also show that if we allow larger metallicities, we obtain $\zeta_{\mathrm{acc,0}} \la 5 \times 10^{-2}$ with $f_{\mathrm{in}}=0.1$ and $\zeta_\mathrm{SN}=0.1$. If we adopt the above star formation time-scale, this corresponds to $\tau_\mathrm{acc,0}\sim 8.5\times 10^6$ yr, which is consistent with the accretion time-scale derived for nearby galaxies \citep{Hirashita:2011aa}.
Note that the dust emission is only tentatively detected for CLM1.
In other words, the detection of dust continuum by ALMA
at $z\ga 6$ indicates that the object should satisfy either of the following conditions: (i) very efficient
condensation in stellar ejecta ($f_\mathrm{in}\ga 0.5$), (ii) very efficient dust growth by accretion
with $\zeta_\mathrm{acc,0}\la 4\times 10^{-3}$, or (iii) high metallicity ($\ga$ solar metallicity).

Recently, \citet{Ferrara:2016aa} proposed another reason for
the `deficit' of dust emission for high-$z$ LBGs. Since the ISM in high-$z$
galaxies has a high pressure, high-$z$ LBGs host dense molecular clouds
\citep{Pallottini:2016aa}.
Thus, a large fraction of dust is `hidden' in the shielded environment of the
dense molecular clouds. Since such a hidden component of dust has
dust temperatures as low as the CMB temperature
at the redshift of the LBGs, its far-infrared emission is weak. If this
deficit of dust emission is common for high-$z$ LBGs, more extreme conditions
(i.e.\ higher efficiencies)
for dust enrichment than shown in this paper would be necessary for
ALMA detection of dust emission.

The above conditions for ALMA detection may be modified if we consider
gas inflow and outflow from galaxies.
We adopted a closed-box model in this paper for simplicity. As mentioned at the beginning
of Section \ref{subsec:enrichment}, if we include the effect of inflow and outflow,
we need to consider the dilution of metals and dust \citep{Feldmann:2015aa}.
However, if we define the total baryonic mass available in the end as $M_0$
and define the star formation time-scale by the total gas mass including
what the galaxy obtain in its entire lifetime, we could use the same framework.
Moreover, we have confirmed that we obtain a similar constraint on
the accretion time-scale to that obtained by \citet{Mancini:2015aa}, who took
into account the build-up of galaxy mass.
Thus, we expect that our conclusion does not change significantly even if we
take into account realistic gas inflow and outflow that occur in the course
of galaxy evolution.

\subsection{Dependence on grain properties}\label{subsec:grain_properties}

Our model assumes dust properties such as the grain radius $a$, the dust material density $s$, the emissivity index $\beta$, and the mass absorption coefficient at 158 $\micron$ $\kappa_{158}$.
Here we investigate how much the constraints obtained for the parameters (such as shown in Fig.\ \ref{fig:constraint}) are affected by the assumed dust properties. Since similar results are obtained for both galaxies, we concentrate on A1689-zD1 in this subsection. We also fix the value of $f_\mathrm{in}$ to 0.1.

\begin{figure}
\begin{center}
\includegraphics[width=0.45\textwidth]{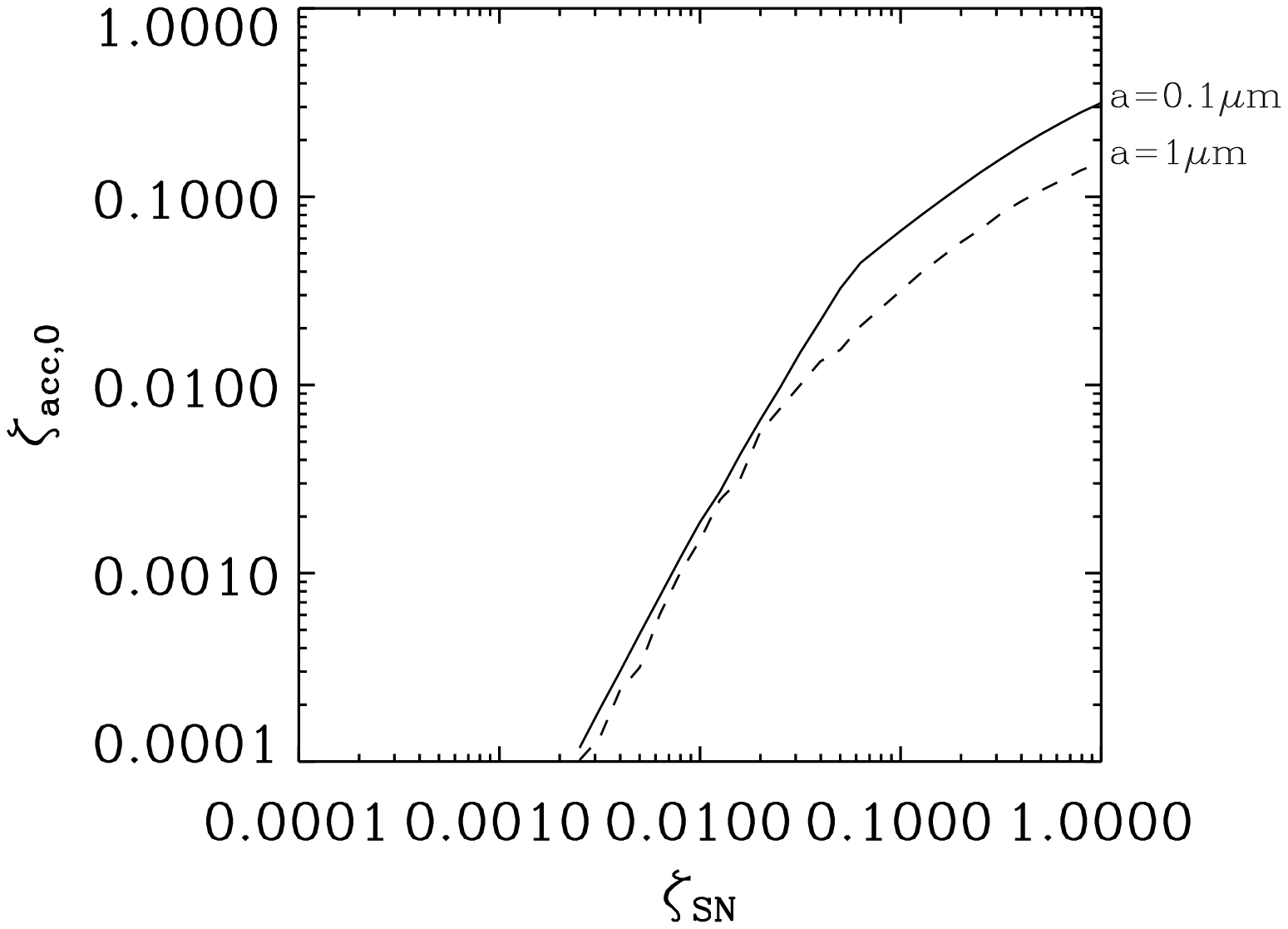}
\includegraphics[width=0.45\textwidth]{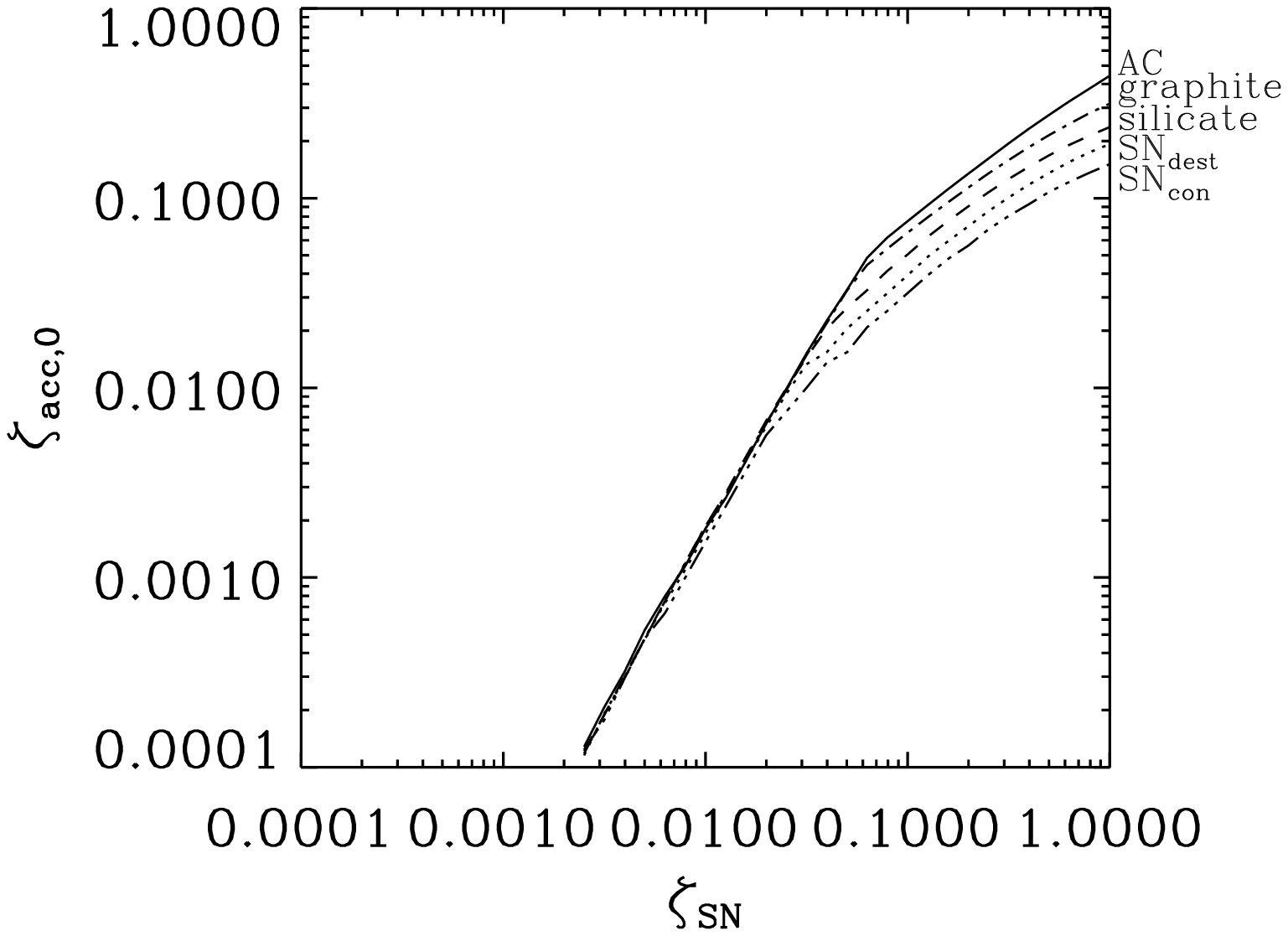}
\caption{Same as Fig.\ \ref{fig:constraint} but for various grain radii and material properties.
Only the boundary for $f_\mathrm{in}=0.1$ is shown. Upper panel: Grain size dependence. The solid and dashed lines refer to $a=0.1$ and 1 $\mu$m, respectively. Lower panel: Dust material dependence.
The lines from top to bottom refer to AC, graphite, silicate, SN$_\mathrm{dest}$, and SN$_\mathrm{con}$, respectively.}
\label{fig:dep_grain}
\end{center}
\end{figure}

A larger grain radius leads to a more efficient emission, thus, a lower temperature (see equation \ref{eq:rad_eq}; the emission scales with $a^3$ while the absorption scales with $a^2$). Thus, we may obtain a severer constraint on the parameters for a larger grain. We here investigate the case of $a=1~\micron$ in Fig.\ \ref{fig:dep_grain}a. We find that the constraints on $\zeta_\mathrm{acc,0}$ and $\zeta_\mathrm{SN}$ differ only by a factor of $\sim 2$ between $a=0.1~\mu$m and 1 $\mu$m.

Variation of the dust material affects the values of $s$, $\beta$, and $\kappa_{158}$. For the dust material, we adopted graphite above, but we here investigate other grain species listed as possible dust species by \citet{Hirashita:2014aa}; silicate, amorphous carbon (AC),  SN$_\mathrm{con}$ (theoretically expected dust material properties for condensation in supernova ejecta; \citet{Nozawa:2003aa}) and SN$_\mathrm{des}$ (theoretically expected values with reverse shock destruction in a supernova remnant; \citet{Nozawa:2007aa}). The dust properties are summarized in Table \ref{tab:kappa}. The constraints obtained for each grain species are compared in Fig.\ \ref{fig:dep_grain}b. The longest accretion time-scale is allowed for AC among all the species, since AC has the highest emissivity, thus achieving the observed ALMA flux most easily. For the opposite reason, SN$_\mathrm{con}$ requires the shortest accretion time-scale.

\begin{table}
\centering
\begin{minipage}{80mm}
\caption{Dust properties.}
\label{tab:kappa}
    \begin{tabular}{lccccccc}
     \hline
     Species & $\kappa_{158}\,^\mathrm{a}$ & $\beta\,^\mathrm{b}$ &
     $s\,^\mathrm{c}$ & Ref.\,$^\mathrm{d}$\\
      & (cm$^2$ g$^{-1}$) & & (g cm$^{-3}$) & \\
     \hline
     Graphite & 20.9 & 2 & 2.26 & 1, 2\\
     Silicate & 13.2 & 2 & 3.3  & 1, 2\\
     SN$_\mathrm{con}\,^\mathrm{e}$ & 5.57 & 1.6 & 2.96 & 3\\
     SN$_\mathrm{dest}\,^\mathrm{f}$ & 8.94 & 2.1 & 2.48 & 4\\
     AC$^\mathrm{g}$ & 28.4 & 1.4 & {1.81}  & 5, 6\\
     \hline
    \end{tabular}
    
    \medskip

$^\mathrm{a}$Mass absorption coefficient at 158 $\micron$.\\
$^\mathrm{b}$Emissivity index ($\kappa_\nu\propto\nu^\beta$).\\
$^\mathrm{c}$Material density.\\
$^\mathrm{d}$ References: 1) Draine \& Lee (1984); 2) Dayal, Hirashita, \& Ferrara (2010);
3) Hirashita et al.\ (2005); 4) Hirashita et al.\ (2008); 5) Zubko et al.\ (1996);
6) Zubko et al.\ (2004).\\
$^\mathrm{e}$Dust condensed in SNe before {reverse} shock destruction
(Nozawa et al.\ 2003).\\
$^\mathrm{f}$Dust ejected from SNe after {reverse} shock destruction
(Nozawa et al.\ 2007).\\
$^\mathrm{g}$Amorphous carbon.
\end{minipage}
\end{table}

\subsection{Importance of metallicity measurements}\label{subsec:metallicity}
In the above, two solutions were broadly permitted to explain the observed
dust abundance in the two high-$z$ galaxies. One is an extremely high
condensation efficiency $f_\mathrm{in}$ typically $\geq 0.5$, although this
solution is not supported by theoretical condensation calculations,
and the other is an efficient accretion. As shown in \citet{Asano:2013ab},
the rise of dust abundance occurs after the system is enriched with
metals to a certain `critical' metallicity, since efficient accretion occurs
only when a sufficient amount of accreting materials (i.e.\ metals) is
available.

Now we reformulate the critical metallicity, which is estimated by
equating the dust formation rates of stellar sources and accretion,
i.e.\ the second and the fourth terms on the right-hand side of equation (\ref{eq:dMd_dt}).
The solution gives the critical metallicity in the sense that, if the
metallicity is higher than this, accretion is the dominant mechanism
of dust production over stellar dust production. As a consequence,
we obtain the critical metallicity as (see Appendix for the derivation)
\begin{equation}
\left(\frac{Z}{\mathrm{Z}_{\sun}}\right)_\mathrm{cr}\simeq\zeta_\mathrm{acc,0}^{1/2},
\end{equation}
where the subscript `cr' indicates that this metallicity gives the
critical metallicity for accretion (see also \citet{Feldmann:2015aa}).

Under the metallicity constraint $Z<0.1$ Z$_{\sun}$, values of $\zeta_\mathrm{acc,0}$ smaller than $4 \times 10^{-3}$ appropriate for the sample of $z>6$ galaxies indicate that the critical metallicity is $\sim 0.06$ Z$_{\sun}$. Therefore, if we find that the metallicity is $0.06 Z_{\sun}<Z<0.1 Z_{\sun}$, it would strongly supports the scenario of efficient dust growth by accretion in those galaxies. Otherwise, for $Z < 0.06 Z_{\sun}$, efficient stellar dust production would be the way to explain the dust emission detected by ALMA. Taking away the metallicity constraint, we obtained a constraint $\zeta_\mathrm{acc,0}\la 0.1$ (i.e. a similar efficiency of accretion to the one in nearby galaxies) for $\zeta_\mathrm{SN}\sim 0.1$, leading to a critical metallicity of $\sim$0.3 Z$_{\sun}$. Thus, if the metallicity is proves to be larger than 0.3 Z$_{\sun}$, accretion naturally explains the high dust luminosity, but it is not necessary to assume an extremely high accretion efficiency compared with the one in nearby galaxies.

It is generally difficult to detect metal emission lines for galaxies at $z \ga 6$. \citet{2012A&A...542L..34N} proposed to use far infrared fine structure lines to assess the metallicity. Future sensitive submm observations may constrain the metallicities in high-z LBGs. Alternatively, the grain size distribution would serve to constrain the dominant dust production process, since the dust grains produced by stars have large sizes ($\approx 0.1$ $\mu$m), while there should be a large abundance of small grains produced by shattering if dust growth by accretion is actively occurring \citep{Asano:2013ab}. Probably, extinction curves, although they are not easy to be derived, could provide the information on the grain size distribution \citep{Asano:2014aa}. Observations of dense molecular gas would directly probe the existence of clouds hosting dust growth by accretion, since extremely dense molecular clouds may enhance the efficiency of dust growth \citep{Kuo:2013aa}. Indeed,  \citet{Pallottini:2016aa} showed in their simulation a dense and metal-enriched region in the central part of an LBG. Such a dense region may host dust growth. Observations of highly excited molecular line whose critical density is high may serve to identify such a dense region \citep{2016arXiv160608464V}. Thus, if observations of dense molecular clouds prove that molecular clouds are dense (by high excitation molecular lines, etc.), they would support the possibility of efficient accretion.

\section{Conclusion}

Recent ALMA observations have started to detect normal galaxies even at $z>6$.
We investigated what constraint we can get for the dust enrichment in
normal $z>6$ galaxies detected by ALMA. To this aim, we used a theoretical model that includes major processes driving dust evolution in a galaxy, including dust condensation in stellar ejecta, dust growth by the accretion of gas-phase metals, and supernova destruction.
To cancel out unknown quantities such as the total baryonic content and the star formation
time-scale, we consider the flux at the ALMA band normalized to the UV flux as a function
of time normalized to the star formation time-scale.
Using the observational data for A1689-zD1 and CLM1, we
obtained the range of the time-scales (or efficiencies) of the above mentioned processes
by examining if the observed level of the ALMA flux normalized to the UV flux is
achieved. We find that if we assume extremely high condensation efficiency in stellar ejecta ($f_{\mathrm{in}} \ga 0.5$), stardust may be enough to explain the observed ALMA flux in their early evolutionary stage, unless dust destruction by supernovae in those galaxies is stronger than that in nearby galaxies.
If we assume a condensation efficiency expected from theoretical calculations ($f_{\mathrm{in}} \la 0.1$), dust growth by accretion is required.
If the metallicities of those objects are lower than 0.1 Z$_{\sun}$ as suggested by
a previous theoretical model \citep{Mancini:2015aa},
strong dust growth (ten times stronger than assumed for nearby galaxies) is derived.
Alternatively, if we allow for solar metallicity, only moderate accretion whose efficiency is comparable to that in nearby galaxies is required to explain the ALMA detections.  
These results indicate that the normal galaxies detected by ALMA at $z>6$ are biased to objects (i) with high dust condensation efficiency in stellar ejecta, (ii) with strong dust growth, or (iii) with efficient dust growth because of fast metal enrichment up to solar metallicity. We finally argue that measurement of metallicity will provide a key to distinguishing among these possibilities. 

\section*{Acknowledgment}
We are grateful to the anonymous referee for useful comments.
HH thanks the Ministry of Science and Technology for support through grant
MOST 105-2112-M-001-027-MY3.

\bibliographystyle{mnras}
\bibliography{Reference}

\begin{thebibliography}{}
\makeatletter
\relax
\def\mn@urlcharsother{\let\do\@makeother \do\$\do\&\do\#\do\^\do\_\do\%\do\~}
\def\mn@doi{\begingroup\mn@urlcharsother \@ifnextchar [ {\mn@doi@}
  {\mn@doi@[]}}
\def\mn@doi@[#1]#2{\def\@tempa{#1}\ifx\@tempa\@empty \href
  {http://dx.doi.org/#2} {doi:#2}\else \href {http://dx.doi.org/#2} {#1}\fi
  \endgroup}
\def\mn@eprint#1#2{\mn@eprint@#1:#2::\@nil}
\def\mn@eprint@arXiv#1{\href {http://arxiv.org/abs/#1} {{\tt arXiv:#1}}}
\def\mn@eprint@dblp#1{\href {http://dblp.uni-trier.de/rec/bibtex/#1.xml}
  {dblp:#1}}
\def\mn@eprint@#1:#2:#3:#4\@nil{\def\@tempa {#1}\def\@tempb {#2}\def\@tempc
  {#3}\ifx \@tempc \@empty \let \@tempc \@tempb \let \@tempb \@tempa \fi \ifx
  \@tempb \@empty \def\@tempb {arXiv}\fi \@ifundefined
  {mn@eprint@\@tempb}{\@tempb:\@tempc}{\expandafter \expandafter \csname
  mn@eprint@\@tempb\endcsname \expandafter{\@tempc}}}

\bibitem[\protect\citeauthoryear{{Aravena} et~al.,}{{Aravena}
  et~al.}{2016}]{Aravena:2016aa}
{Aravena} M.,  et~al., 2016, preprint

\bibitem[\protect\citeauthoryear{{Asano}, {Takeuchi}, {Hirashita}  \&
  {Inoue}}{{Asano} et~al.}{2013}]{Asano:2013ab}
{Asano} R.~S.,  {Takeuchi} T.~T.,  {Hirashita} H.,   {Inoue} A.~K.,  2013,
  \mn@doi [Earth, Planets, and Space] {10.5047/eps.2012.04.014}, \href
  {http://adsabs.harvard.edu/abs/2013EP%26S...65..213A} {65, 213}

\bibitem[\protect\citeauthoryear{{Asano}, {Takeuchi}, {Hirashita}  \&
  {Nozawa}}{{Asano} et~al.}{2014}]{Asano:2014aa}
{Asano} R.~S.,  {Takeuchi} T.~T.,  {Hirashita} H.,   {Nozawa} T.,  2014,
  \mn@doi [\mnras] {10.1093/mnras/stu208}, \href
  {http://adsabs.harvard.edu/abs/2014MNRAS.440..134A} {440, 134}

\bibitem[\protect\citeauthoryear{{Bianchi} \& {Schneider}}{{Bianchi} \&
  {Schneider}}{2007}]{Bianchi:2007aa}
{Bianchi} S.,  {Schneider} R.,  2007, \mn@doi [\mnras]
  {10.1111/j.1365-2966.2007.11829.x}, \href
  {http://adsabs.harvard.edu/abs/2007MNRAS.378..973B} {378, 973}

\bibitem[\protect\citeauthoryear{{Bouwens} et~al.,}{{Bouwens}
  et~al.}{2016}]{Bouwens:2016aa}
{Bouwens} R.,  et~al., 2016, preprint, \href
  {http://adsabs.harvard.edu/abs/2016arXiv160605280B} {} (\mn@eprint {arXiv}
  {1606.05280})

\bibitem[\protect\citeauthoryear{{Capak} et~al.,}{{Capak}
  et~al.}{2015}]{Capak:2015aa}
{Capak} P.~L.,  et~al., 2015, \mn@doi [\nat] {10.1038/nature14500}, \href
  {http://adsabs.harvard.edu/abs/2015Natur.522..455C} {522, 455}

\bibitem[\protect\citeauthoryear{{Carroll}, {Press}  \& {Turner}}{{Carroll}
  et~al.}{1992}]{Carroll:1992aa}
{Carroll} S.~M.,  {Press} W.~H.,   {Turner} E.~L.,  1992, \mn@doi [\araa]
  {10.1146/annurev.aa.30.090192.002435}, \href
  {http://adsabs.harvard.edu/abs/1992ARA%26A..30..499C} {30, 499}

\bibitem[\protect\citeauthoryear{{\lowercase{D}a Cunha}
  et~al.,}{{\lowercase{D}a Cunha} et~al.}{2013}]{da-Cunha:2013aa}
{\lowercase{D}a Cunha} E.,  et~al., 2013, \mn@doi [\apj]
  {10.1088/0004-637X/766/1/13}, \href
  {http://adsabs.harvard.edu/abs/2013ApJ...766...13D} {766, 13}

\bibitem[\protect\citeauthoryear{{Dayal}, {Hirashita}  \& {Ferrara}}{{Dayal}
  et~al.}{2010}]{Dayal:2010aa}
{Dayal} P.,  {Hirashita} H.,   {Ferrara} A.,  2010, \mn@doi [\mnras]
  {10.1111/j.1365-2966.2009.16164.x}, \href
  {http://adsabs.harvard.edu/abs/2010MNRAS.403..620D} {403, 620}

\bibitem[\protect\citeauthoryear{{\lowercase{D}e Bennassuti}, {Schneider},
  {Valiante}  \& {Salvadori}}{{\lowercase{D}e Bennassuti}
  et~al.}{2014}]{de-Bennassuti:2014aa}
{\lowercase{D}e Bennassuti} M.,  {Schneider} R.,  {Valiante} R.,   {Salvadori}
  S.,  2014, \mn@doi [\mnras] {10.1093/mnras/stu1962}, \href
  {http://adsabs.harvard.edu/abs/2014MNRAS.445.3039D} {445, 3039}

\bibitem[\protect\citeauthoryear{{Dwek}}{{Dwek}}{1998}]{Dwek:1998aa}
{Dwek} E.,  1998, \mn@doi [\apj] {10.1086/305829}, \href
  {http://adsabs.harvard.edu/abs/1998ApJ...501..643D} {501, 643}

\bibitem[\protect\citeauthoryear{{Dwek} \& {Scalo}}{{Dwek} \&
  {Scalo}}{1980}]{Dwek:1980aa}
{Dwek} E.,  {Scalo} J.~M.,  1980, \mn@doi [\apj] {10.1086/158100}, \href
  {http://adsabs.harvard.edu/abs/1980ApJ...239..193D} {239, 193}

\bibitem[\protect\citeauthoryear{{Feldmann}}{{Feldmann}}{2015}]{Feldmann:2015aa}
{Feldmann} R.,  2015, \mn@doi [\mnras] {10.1093/mnras/stv552}, \href
  {http://adsabs.harvard.edu/abs/2015MNRAS.449.3274F} {449, 3274}

\bibitem[\protect\citeauthoryear{{Ferrara}, {Hirashita}, {Ouchi}  \&
  {Fujimoto}}{{Ferrara} et~al.}{2016}]{Ferrara:2016aa}
{Ferrara} A.,  {Hirashita} H.,  {Ouchi} M.,   {Fujimoto} S.,  2016, preprint

\bibitem[\protect\citeauthoryear{{Gall}, {Andersen}  \& {Hjorth}}{{Gall}
  et~al.}{2011}]{Gall:2011aa}
{Gall} C.,  {Andersen} A.~C.,   {Hjorth} J.,  2011, \mn@doi [\aap]
  {10.1051/0004-6361/201015286}, \href
  {http://adsabs.harvard.edu/abs/2011A%26A...528A..13G} {528, A13}

\bibitem[\protect\citeauthoryear{{Gould} \& {Salpeter}}{{Gould} \&
  {Salpeter}}{1963}]{Gould:1963ab}
{Gould} R.~J.,  {Salpeter} E.~E.,  1963, \mn@doi [\apj] {10.1086/147654}, \href
  {http://adsabs.harvard.edu/abs/1963ApJ...138..393G} {138, 393}

\bibitem[\protect\citeauthoryear{{Hirashita}}{{Hirashita}}{1999}]{Hirashita:1999aa}
{Hirashita} H.,  1999, \mn@doi [\apjl] {10.1086/311806}, \href
  {http://adsabs.harvard.edu/abs/1999ApJ...510L..99H} {510, L99}

\bibitem[\protect\citeauthoryear{{Hirashita} \& {Kuo}}{{Hirashita} \&
  {Kuo}}{2011}]{Hirashita:2011aa}
{Hirashita} H.,  {Kuo} T.-M.,  2011, \mn@doi [\mnras]
  {10.1111/j.1365-2966.2011.19131.x}, \href
  {http://adsabs.harvard.edu/abs/2011MNRAS.416.1340H} {416, 1340}

\bibitem[\protect\citeauthoryear{{Hirashita}, {Ferrara}, {Dayal}  \&
  {Ouchi}}{{Hirashita} et~al.}{2014}]{Hirashita:2014aa}
{Hirashita} H.,  {Ferrara} A.,  {Dayal} P.,   {Ouchi} M.,  2014, \mn@doi
  [\mnras] {10.1093/mnras/stu1290}, \href
  {http://adsabs.harvard.edu/abs/2014MNRAS.443.1704H} {443, 1704}

\bibitem[\protect\citeauthoryear{{Inoue}}{{Inoue}}{2003}]{Inoue:2003aa}
{Inoue} A.~K.,  2003, \mn@doi [\pasj] {10.1093/pasj/55.5.901}, \href
  {http://adsabs.harvard.edu/abs/2003PASJ...55..901I} {55, 901}

\bibitem[\protect\citeauthoryear{{Inoue}}{{Inoue}}{2011}]{Inoue:2011aa}
{Inoue} A.~K.,  2011, \mn@doi [Earth, Planets, and Space]
  {10.5047/eps.2011.02.013}, \href
  {http://adsabs.harvard.edu/abs/2011EP%26S...63.1027I} {63, 1027}

\bibitem[\protect\citeauthoryear{{Knudsen}, {Watson}, {Frayer}, {Christensen},
  {Gallazzi}, {Michalowski}, {Richard}  \& {Zavala}}{{Knudsen}
  et~al.}{2016}]{Knudsen:2016aa}
{Knudsen} K.~K.,  {Watson} D.,  {Frayer} D.,  {Christensen} L.,  {Gallazzi} A.,
   {Michalowski} M.~J.,  {Richard} J.,   {Zavala} J.,  2016, preprint, \href
  {http://adsabs.harvard.edu/abs/2016arXiv160303222K} {} (\mn@eprint {arXiv}
  {1603.03222})

\bibitem[\protect\citeauthoryear{{Kozasa}, {Hasegawa}  \& {Nomoto}}{{Kozasa}
  et~al.}{1989}]{Kozasa:1989aa}
{Kozasa} T.,  {Hasegawa} H.,   {Nomoto} K.,  1989, \mn@doi [\apj]
  {10.1086/167801}, \href {http://adsabs.harvard.edu/abs/1989ApJ...344..325K}
  {344, 325}

\bibitem[\protect\citeauthoryear{{Kuo}, {Hirashita}  \& {Zafar}}{{Kuo}
  et~al.}{2013}]{Kuo:2013aa}
{Kuo} T.-M.,  {Hirashita} H.,   {Zafar} T.,  2013, \mn@doi [\mnras]
  {10.1093/mnras/stt1648}, \href
  {http://adsabs.harvard.edu/abs/2013MNRAS.436.1238K} {436, 1238}

\bibitem[\protect\citeauthoryear{{Lisenfeld} \& {Ferrara}}{{Lisenfeld} \&
  {Ferrara}}{1998}]{Lisenfeld:1998aa}
{Lisenfeld} U.,  {Ferrara} A.,  1998, \mn@doi [\apj] {10.1086/305354}, \href
  {http://adsabs.harvard.edu/abs/1998ApJ...496..145L} {496, 145}

\bibitem[\protect\citeauthoryear{{Maiolino} et~al.,}{{Maiolino}
  et~al.}{2015}]{Maiolino:2015aa}
{Maiolino} R.,  et~al., 2015, \mn@doi [\mnras] {10.1093/mnras/stv1194}, \href
  {http://adsabs.harvard.edu/abs/2015MNRAS.452...54M} {452, 54}

\bibitem[\protect\citeauthoryear{{Mancini}, {Schneider}, {Graziani},
  {Valiante}, {Dayal}, {Maio}, {Ciardi}  \& {Hunt}}{{Mancini}
  et~al.}{2015}]{Mancini:2015aa}
{Mancini} M.,  {Schneider} R.,  {Graziani} L.,  {Valiante} R.,  {Dayal} P.,
  {Maio} U.,  {Ciardi} B.,   {Hunt} L.~K.,  2015, \mn@doi [\mnras]
  {10.1093/mnrasl/slv070}, \href
  {http://adsabs.harvard.edu/abs/2015MNRAS.451L..70M} {451, L70}

\bibitem[\protect\citeauthoryear{{Mattsson} \& {Andersen}}{{Mattsson} \&
  {Andersen}}{2012}]{Mattsson:2012aa}
{Mattsson} L.,  {Andersen} A.~C.,  2012, \mn@doi [\mnras]
  {10.1111/j.1365-2966.2012.20574.x}, \href
  {http://adsabs.harvard.edu/abs/2012MNRAS.423...38M} {423, 38}

\bibitem[\protect\citeauthoryear{{McKee}}{{McKee}}{1989}]{McKee:1989aa}
{McKee} C.,  1989, in {Allamandola} L.~J.,  {Tielens} A.~G.~G.~M.,  eds,  IAU
  Symposium Vol. 135, Interstellar Dust. p.~431

\bibitem[\protect\citeauthoryear{{Micha{\l}owski}}{{Micha{\l}owski}}{2015}]{Michalowski:2015aa}
{Micha{\l}owski} M.~J.,  2015, \mn@doi [\aap] {10.1051/0004-6361/201525644},
  \href {http://adsabs.harvard.edu/abs/2015A%26A...577A..80M} {577, A80}

\bibitem[\protect\citeauthoryear{{Nagao}, {Maiolino}, {De Breuck}, {Caselli},
  {Hatsukade}  \& {Saigo}}{{Nagao} et~al.}{2012}]{2012A&A...542L..34N}
{Nagao} T.,  {Maiolino} R.,  {De Breuck} C.,  {Caselli} P.,  {Hatsukade} B.,
  {Saigo} K.,  2012, \mn@doi [\aap] {10.1051/0004-6361/201219518}, \href
  {http://ads.nao.ac.jp/abs/2012A%26A...542L..34N} {542, L34}

\bibitem[\protect\citeauthoryear{{Nozawa}, {Kozasa}, {Umeda}, {Maeda}  \&
  {Nomoto}}{{Nozawa} et~al.}{2003}]{Nozawa:2003aa}
{Nozawa} T.,  {Kozasa} T.,  {Umeda} H.,  {Maeda} K.,   {Nomoto} K.,  2003,
  \mn@doi [\apj] {10.1086/379011}, \href
  {http://adsabs.harvard.edu/abs/2003ApJ...598..785N} {598, 785}

\bibitem[\protect\citeauthoryear{{Nozawa}, {Kozasa}  \& {Habe}}{{Nozawa}
  et~al.}{2006}]{Nozawa:2006aa}
{Nozawa} T.,  {Kozasa} T.,   {Habe} A.,  2006, \mn@doi [\apj] {10.1086/505639},
  \href {http://adsabs.harvard.edu/abs/2006ApJ...648..435N} {648, 435}

\bibitem[\protect\citeauthoryear{{Nozawa}, {Kozasa}, {Habe}, {Dwek}, {Umeda},
  {Tominaga}, {Maeda}  \& {Nomoto}}{{Nozawa} et~al.}{2007}]{Nozawa:2007aa}
{Nozawa} T.,  {Kozasa} T.,  {Habe} A.,  {Dwek} E.,  {Umeda} H.,  {Tominaga} N.,
   {Maeda} K.,   {Nomoto} K.,  2007, \mn@doi [\apj] {10.1086/520621}, \href
  {http://adsabs.harvard.edu/abs/2007ApJ...666..955N} {666, 955}

\bibitem[\protect\citeauthoryear{{Ota} et~al.,}{{Ota}
  et~al.}{2014}]{Ota:2014aa}
{Ota} K.,  et~al., 2014, \mn@doi [\apj] {10.1088/0004-637X/792/1/34}, \href
  {http://adsabs.harvard.edu/abs/2014ApJ...792...34O} {792, 34}

\bibitem[\protect\citeauthoryear{{Ouchi} et~al.,}{{Ouchi}
  et~al.}{2013}]{Ouchi:2013aa}
{Ouchi} M.,  et~al., 2013, \mn@doi [\apj] {10.1088/0004-637X/778/2/102}, \href
  {http://adsabs.harvard.edu/abs/2013ApJ...778..102O} {778, 102}

\bibitem[\protect\citeauthoryear{{Pallottini}, {Ferrara}, {Gallerani},
  {Vallini}, {Maiolino}  \& {Salvadori}}{{Pallottini}
  et~al.}{2016}]{Pallottini:2016aa}
{Pallottini} A.,  {Ferrara} A.,  {Gallerani} S.,  {Vallini} L.,  {Maiolino} R.,
    {Salvadori} S.,  2016, preprint, \href
  {http://adsabs.harvard.edu/abs/2016arXiv160901719P} {} (\mn@eprint {arXiv}
  {1609.01719})

\bibitem[\protect\citeauthoryear{{R{\'e}my-Ruyer} et~al.,}{{R{\'e}my-Ruyer}
  et~al.}{2014}]{Remy-Ruyer:2014aa}
{R{\'e}my-Ruyer} A.,  et~al., 2014, \mn@doi [\aap]
  {10.1051/0004-6361/201322803}, \href
  {http://adsabs.harvard.edu/abs/2014A%26A...563A..31R} {563, A31}

\bibitem[\protect\citeauthoryear{{Schaerer}, {Boone}, {Zamojski}, {Staguhn},
  {Dessauges-Zavadsky}, {Finkelstein}  \& {Combes}}{{Schaerer}
  et~al.}{2015}]{Schaerer:2015aa}
{Schaerer} D.,  {Boone} F.,  {Zamojski} M.,  {Staguhn} J.,
  {Dessauges-Zavadsky} M.,  {Finkelstein} S.,   {Combes} F.,  2015, \mn@doi
  [\aap] {10.1051/0004-6361/201424649}, \href
  {http://adsabs.harvard.edu/abs/2015A%26A...574A..19S} {574, A19}

\bibitem[\protect\citeauthoryear{{Schneider}, {Hunt}  \&
  {Valiante}}{{Schneider} et~al.}{2016}]{Schneider:2016aa}
{Schneider} R.,  {Hunt} L.,   {Valiante} R.,  2016, \mn@doi [\mnras]
  {10.1093/mnras/stw114}, \href
  {http://adsabs.harvard.edu/abs/2016MNRAS.457.1842S} {457, 1842}

\bibitem[\protect\citeauthoryear{{Tinsley}}{{Tinsley}}{1980}]{Tinsley:1980aa}
{Tinsley} B.~M.,  1980, \fcp, \href
  {http://adsabs.harvard.edu/abs/1980FCPh....5..287T} {5, 287}

\bibitem[\protect\citeauthoryear{{Todini} \& {Ferrara}}{{Todini} \&
  {Ferrara}}{2001}]{Todini:2001aa}
{Todini} P.,  {Ferrara} A.,  2001, \mn@doi [\mnras]
  {10.1046/j.1365-8711.2001.04486.x}, \href
  {http://adsabs.harvard.edu/abs/2001MNRAS.325..726T} {325, 726}

\bibitem[\protect\citeauthoryear{{Valiante}, {Schneider}, {Salvadori}  \&
  {Bianchi}}{{Valiante} et~al.}{2011}]{Valiante:2011aa}
{Valiante} R.,  {Schneider} R.,  {Salvadori} S.,   {Bianchi} S.,  2011, \mn@doi
  [\mnras] {10.1111/j.1365-2966.2011.19168.x}, \href
  {http://adsabs.harvard.edu/abs/2011MNRAS.416.1916V} {416, 1916}

\bibitem[\protect\citeauthoryear{{Vallini}, {Ferrara}, {Pallottini}  \&
  {Gallerani}}{{Vallini} et~al.}{2016}]{2016arXiv160608464V}
{Vallini} L.,  {Ferrara} A.,  {Pallottini} A.,   {Gallerani} S.,  2016,
  preprint, \href {http://ads.nao.ac.jp/abs/2016arXiv160608464V} {} (\mn@eprint
  {arXiv} {1606.08464})

\bibitem[\protect\citeauthoryear{{Watson}, {Christensen}, {Knudsen}, {Richard},
  {Gallazzi}  \& {Micha{\l}owski}}{{Watson} et~al.}{2015}]{Watson:2015aa}
{Watson} D.,  {Christensen} L.,  {Knudsen} K.~K.,  {Richard} J.,  {Gallazzi}
  A.,   {Micha{\l}owski} M.~J.,  2015, \mn@doi [\nat] {10.1038/nature14164},
  \href {http://adsabs.harvard.edu/abs/2015Natur.519..327W} {519, 327}

\bibitem[\protect\citeauthoryear{{Willott}, {Carilli}, {Wagg}  \&
  {Wang}}{{Willott} et~al.}{2015}]{Willott:2015aa}
{Willott} C.~J.,  {Carilli} C.~L.,  {Wagg} J.,   {Wang} R.,  2015, \mn@doi
  [\apj] {10.1088/0004-637X/807/2/180}, \href
  {http://adsabs.harvard.edu/abs/2015ApJ...807..180W} {807, 180}

\bibitem[\protect\citeauthoryear{{Yajima}, {Shlosman}, {Romano-D{\'{\i}}az}  \&
  {Nagamine}}{{Yajima} et~al.}{2015}]{Yajima:2015aa}
{Yajima} H.,  {Shlosman} I.,  {Romano-D{\'{\i}}az} E.,   {Nagamine} K.,  2015,
  \mn@doi [\mnras] {10.1093/mnras/stv974}, \href
  {http://adsabs.harvard.edu/abs/2015MNRAS.451..418Y} {451, 418}

\bibitem[\protect\citeauthoryear{{Zhukovska}, {Gail}  \&
  {Trieloff}}{{Zhukovska} et~al.}{2008}]{Zhukovska:2008aa}
{Zhukovska} S.,  {Gail} H.-P.,   {Trieloff} M.,  2008, \mn@doi [\aap]
  {10.1051/0004-6361:20077789}, \href
  {http://adsabs.harvard.edu/abs/2008A%26A...479..453Z} {479, 453}

\makeatother
\end{thebibliography}

\appendix

\section{Critical metallicity for dust growth by accretion}

We compare the two dust-supplying mechanisms, dust formation
by stellar sources and dust growth by accretion. The rates of these
two dust formation paths are represented by the second and
the fourth terms in equation (\ref{eq:dMd_dt}). Because of the dependence
on metallicity, dust growth by accretion becomes dominated over
stellar dust production at a certain metallicity referred to as the
critical metallicity. This critical metallicity is obtained by equating
the above two terms as
\begin{equation}
f_\mathrm{in}\mathcal{Y}_Z\psi\sim\frac{M_\mathrm{d}}{\tau_\mathrm{acc,0}}
\frac{Z}{\mathrm{Z}_{\sun}},\label{eq:stellar_accretion}
\end{equation}
where we adopted $1-f_Z\sim 1$, which holds unless $f_\mathrm{in}$
is nearly unity, for simplicity.
We further use some basic relations,
$\tau_\mathrm{SF}=M_\mathrm{g}/\psi$, $\mathcal{D}=M_\mathrm{d}/M_\mathrm{g}$,
and $\zeta_\mathrm{acc,0}=\tau_\mathrm{acc,0}/\tau_\mathrm{SF}$,
and apply the solution of pure stellar dust production, $\mathcal{D}\sim f_\mathrm{in}Z$,
which is still approximately applicable when accretion starts to be dominant over
stellar dust production. We also use the fact that the metal yield has a similar
value to the solar metallicity (e.g.\ Tinsley 1980). After all,
equation (\ref{eq:stellar_accretion}) is reduced to
\begin{equation}
\left(\frac{Z}{\mathrm{Z}_{\sun}}\right)\sim \zeta_\mathrm{acc,0}^{1/2}.
\end{equation}
This metallicity is referred to as the critical metallicity in Section \ref{subsec:metallicity}.

\bsp

\end{document}